\newif\ifconfver
 \confvertrue        

\newif\ifplainver  
\plainvertrue
 \plainverfalse

\ifplainver
    \confverfalse   
\fi

\ifconfver
     \documentclass[10pt,twocolumn,twoside]{IEEEtran}
\else
    \ifplainver
        \documentclass[11pt]{article}
        \usepackage{fullpage}
    \else
        \documentclass[11pt,draftclsnofoot,onecolumn]{IEEEtran}
    \fi
\fi

\usepackage{graphicx,algorithm,algorithmic,bm,amsmath,amsthm,amssymb,color,shortcuts_OPT,hyperref,cite,tcolorbox,verbatimbox}

\hypersetup{
  colorlinks   = true, 
  urlcolor     = blue, 
  linkcolor    = blue, 
  citecolor    = red   
}

\newtheorem{Def}{Definition}

\usepackage{pgfplots}
\usetikzlibrary{arrows,shapes,calc,tikzmark,backgrounds,matrix,decorations.markings}
\usepgfplotslibrary{fillbetween}

\pgfplotsset{compat=1.3}

\usepackage{relsize}
\tikzset{fontscale/.style = {font=\relsize{#1}}
    }

\definecolor{lavander}{cmyk}{0,0.48,0,0}
\definecolor{violet}{cmyk}{0.79,0.88,0,0}
\definecolor{burntorange}{cmyk}{0,0.52,1,0}

\definecolor{asuorange}{rgb}{1,0.699,0.0625}
\definecolor{asured}{rgb}{0.598,0,0.199}
\definecolor{asuborder}{rgb}{0.953,0.484,0}
\definecolor{asugrey}{rgb}{0.309,0.332,0.340}
\definecolor{asublue}{rgb}{0,0.555,0.836}
\definecolor{asugold}{rgb}{1,0.777,0.008}

\providecommand{\abs}[1]{\lvert#1\rvert} \providecommand{\norm}[1]{\lVert#1\rVert}
    \makeatletter
    \def\multilimits@{\bgroup
  \Let@
  \restore@math@cr
  \default@tag
 \baselineskip\fontdimen10 \scriptfont\tw@
 \advance\baselineskip\fontdimen12 \scriptfont\tw@
 \lineskip\thr@@\fontdimen8 \scriptfont\thr@@
 \lineskiplimit\lineskip
 \vbox\bgroup\ialign\bgroup\hfil$\m@th\scriptstyle{##}$\hfil\crcr}
    \def\Sb{_\multilimits@}
    \def\endSb{\crcr\egroup\egroup\egroup}
\makeatother

\newtheoremstyle{t}         
    {\baselineskip}{2\topsep}      
    {\rm}                   
    {0pt}{\bfseries}  
    {}                      
    { }                      
    {\thmname{#1}\thmnumber{#2}.}

\theoremstyle{t}

\makeatletter
\DeclareRobustCommand*\cal{\@fontswitch\relax\mathcal}
\makeatother

\makeatletter
\renewcommand\paragraph{\@startsection{paragraph}{4}{-\parindent}{1ex}{-.25em}{\normalfont\normalsize\bfseries}}
\makeatother

\begin{document}
\bstctlcite{IEEEexample:BSTcontrol} 
\title{A User Guide to Low-Pass Graph Signal Processing and its Applications}
\author{Raksha Ramakrishna, Hoi-To Wai, Anna Scaglione\thanks{RR, AS are with the School of Electrical, Computer, and Energy Engineering, Arizona State University, AZ, USA. Emails: \texttt{\{rramakr6,Anna.Scaglione\}@asu.edu}. HTW is with the Department of Systems Engineering and Engineering Management, The Chinese University of Hong Kong, Hong Kong SAR of China. Email: \texttt{htwai@se.cuhk.edu.hk}.}}
\date{\today}

\maketitle\vspace{-1.5cm}

\begin{abstract}
The notion of graph filters can be used to define generative models for graph data. In fact, the data obtained from many examples of network dynamics may be viewed as the output of a graph filter. With this interpretation, classical signal processing tools such as frequency analysis have been successfully applied with analogous interpretation to graph data, generating new insights for data science. What follows is a user guide on a specific class of graph data, where the generating graph filters are low-pass, i.e., the filter attenuates contents in the higher graph frequencies while retaining contents in the lower frequencies. Our choice is motivated by the prevalence of low-pass models in application domains such as social networks, financial markets, and power systems. We illustrate how to leverage properties of low-pass graph filters to learn the graph topology or identify its community structure; efficiently represent graph data through sampling, recover missing measurements, and de-noise graph data; the low-pass property is also used as the baseline to detect anomalies.
\end{abstract}
\vspace{-0.2cm}
\section{Introduction} 
A growing trend in signal processing and machine learning is to develop theories and models for analyzing data defined on irregular domains such as graphs.
Graphs often express relational ties, such as social, economics networks, or gene networks, for which several mathematical and statistical models relying on graphs have been proposed to explain trends in the data  \cite{kolaczyk2014statistical}. Another case is that of physical infrastructures (utility networks such as power, gas, water delivery systems and transportation networks) where physical laws, in addition to the connectivity, define the  structure in signals.

For a period of time, the graphical interpretation was primarily used in statistics with the aim of making \emph{inference} about graphical models. Meanwhile, the need for \emph{processing} graph data has led to the emerging field of graph signal processing (GSP), which takes a deterministic and system theoretic approach to justify the  properties of graph data and to inspire the associated signal processing algorithms. A cornerstone of GSP is the formal definition of graph filter,
which extends the notions of linear time invariant (LTI) filtering of time series signals to data defined on a graph, a.k.a.~graph signals.  

\begin{figure*}
    \centering
    \includegraphics[width=0.8\textwidth]{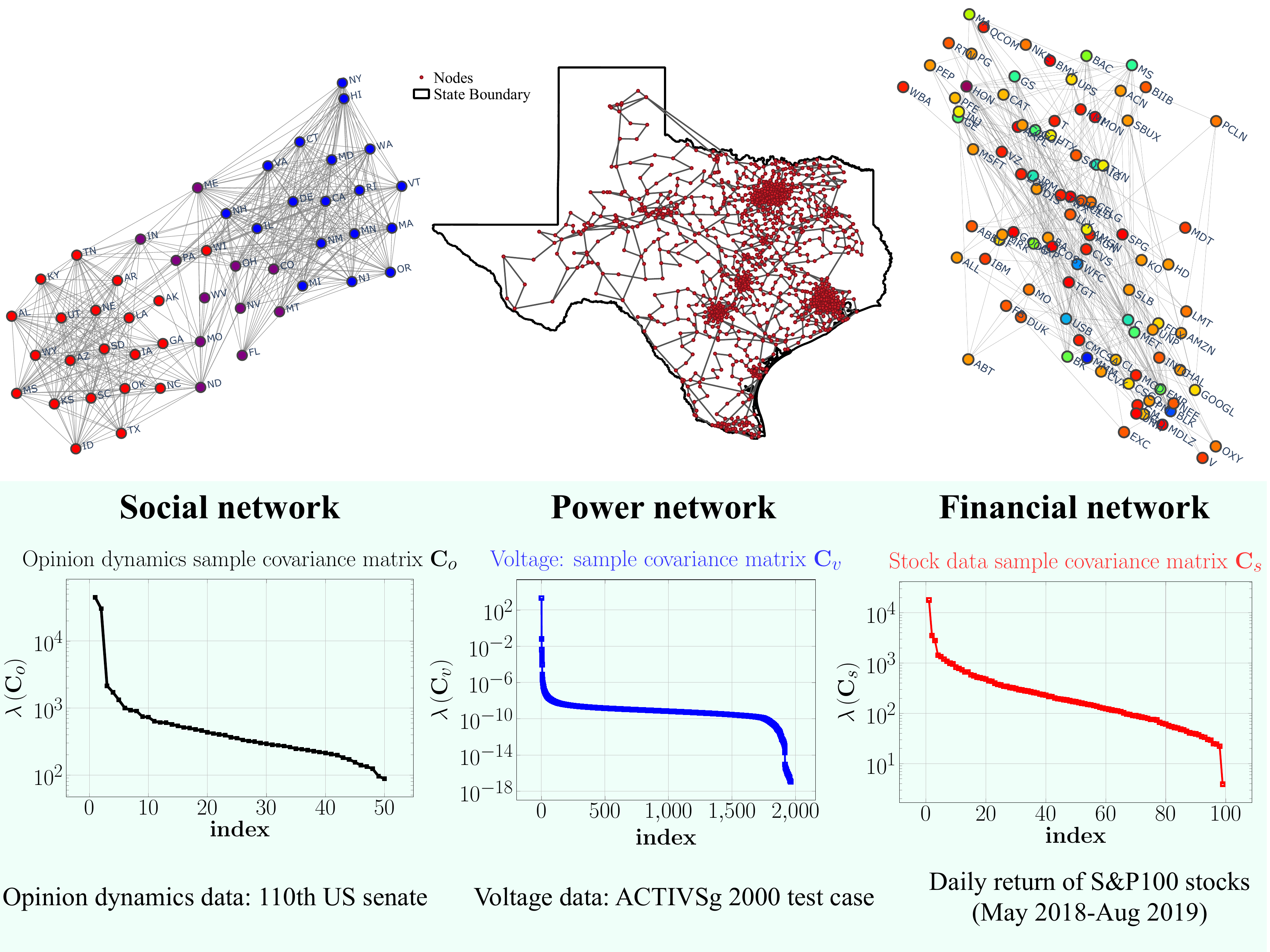}\vspace{-.1cm}
    \caption{Illustrating the eigenvalue spectra of sample data covariance matrix of voltage, Senate rollcall, and financial stock data. These data admit  physical/social models that can be regarded as low-pass filtered graph signals.
    A salient feature of their low-pass nature is observed as the low-rank property of the sample covariance matrices.}\vspace{-.2cm}
    \label{fig:overview_GSP}
\end{figure*}

In a similar vein as LTI filters in discrete-time signal processing, a graph filter can be classified as either low-pass, band-pass, or high-pass, depending on its graph frequency response. 
Among them, this article focuses on the \emph{low-pass} graph filters and the \emph{low-pass} graph signals generated from them. These graph filters capture a smoothing operation applied to the input graph signals, which is a common property of processes observed in many physical/social systems (see Section~\ref{sec:model}).
As a motivating example, in Fig.~\ref{fig:overview_GSP}, we illustrate a few real datasets with such models from social networks, power systems and financial market, and show the eigenvalue spectra of their sample covariance matrices. 
A salient feature observed is that these sample covariance matrices are low-rank, thus displaying an important symptom of low-pass filtered graph signals (to be discussed in Section~\ref{sec:tools}).

Previous articles such as \cite{shuman2013emerging,ortega2018graph} have provided a comprehensive introduction to modeling and processing graph or network data using GSP,  favoring general abstractions over focusing on particular structures and concrete applications.
This \emph{user guide} takes a different approach, concentrating on \emph{low-pass graph filters} and the corresponding \emph{low-pass graph signal} outputs. Low-pass graph signals have specific properties that affect their structure and dictate how to approach, for example, sampling, denoising and inference problems. They are worth focusing on, because they are very common in practice. We start the article by surveying \emph{low-pass} GSP properties and insights, setting the stage for the description of the concrete situations where such a model applies.
A set of particular examples is then provided, highlighting the fact that low-pass graph signals often appear  in different application domains. 
In fact, resorting to existing underlying network dynamical models that justify different data sets, we show that low-pass graph processes are nearly ubiquitous in contexts where GSP is applicable.

\vspace{-.1cm}
\section{Basics of Graph Signal Processing}\label{sec:basic}\vspace{-.1cm}
Many tools introduced in this user guide involve several fundamental concepts of GSP, including a formal definition of low-pass graph filters/signals. These ideas will be briefly reviewed in this section.
For more details, the readers are  referred to the excellent prior overview articles such as \cite{ortega2018graph,shuman2013emerging}. 
We denote vectors with boldfaced lowercase letters, $\bm{x}$ and uppercase letters for matrices, $\mathbf{A}$. The operation $\texttt{Diag}(\bm{x})$ creates a diagonal matrix with elements from vector  $\bm{x}$.

We focus on a weighted undirected graph ${\cal G} = ( {\cal N}, {\cal E} )$ with $n$ nodes such that ${\cal N} = \{1,\ldots,n \}$ and ${\cal E} \subseteq {\cal N} \times {\cal N}$ is the edge set.
A graph signal is a function $x : {\cal N} \rightarrow \RR$ which can be represented by a $n$-dimensional vector  ${\bm x} = (x (i) )_{i \in {\cal N}}$.  
A Graph Shift Operator (GSO) is a matrix $\bm{S} \in \RR^{n \times n}$ satisfying $[\bm{S}]_{ij} \neq 0$ if and only if $i=j$ or $(i,j) \in {\mathcal{E}}$. When multiplied by a graph signal ${\bm x}$, each entry of the shifted graph signal is a linear combination of the one-hop neighbors' values, therefore `shifting' the graph signal with respect to the graph topology.
In this article, we take the Laplacian matrix as the GSO. The Laplacian matrix is defined as ${\bm L} := {\bm D} - {\bm A}$, where ${\bm A}$ is the weighted  symmetric adjacency matrix of ${\cal G}$, and ${\bm D} = \texttt{Diag}( {\bm A} {\bf 1} )$ is a diagonal matrix of the weighted degrees. 
It is also common to take the GSO as the normalized Laplacian matrix, or the adjacency matrix  \cite{sandryhaila2013discrete}.

Having defined the GSO, we discuss how to measure the smoothness of graph signals and analyze their content in the graph frequency domain.
Recall that if a signal is smooth in time, the norm of its time derivative is small. 
For a graph signal ${\bm x}$, its \emph{graph derivative} is defined as
\[
[\grd {\bm x}]_{ij} = \sqrt{A_{ij}} (x_i - x_j).
\]
The squared Frobenius norm of graph derivative, a.k.a.~the \emph{graph quadratic form}  \cite{shuman2013emerging}, provides an idea of the smoothness of the graph signal ${\bm x}$:
\begin{equation}
{\rm S}_{2}( {\bm x} ) := \frac{1}{2} \| \grd {\bm x} \|_{\rm F}^2 = {\bm x}^\top {\bm L} {\bm x} = \sum_{i,j} A_{ij} (x_i-x_j)^2.
\label{eq:total variation}
\end{equation}
Observe that if $x_i \approx x_j$ for any neighboring nodes $i,j$, then ${\rm S}_2( {\bm x}) \approx 0$. As such, we say that a graph signal is \emph{smooth} if ${\rm S}_{2}( {\bm x} ) / \|\bm x\|_2$ is small.

\begin{figure*}
    \centering
    \includegraphics[width=0.7\textwidth]{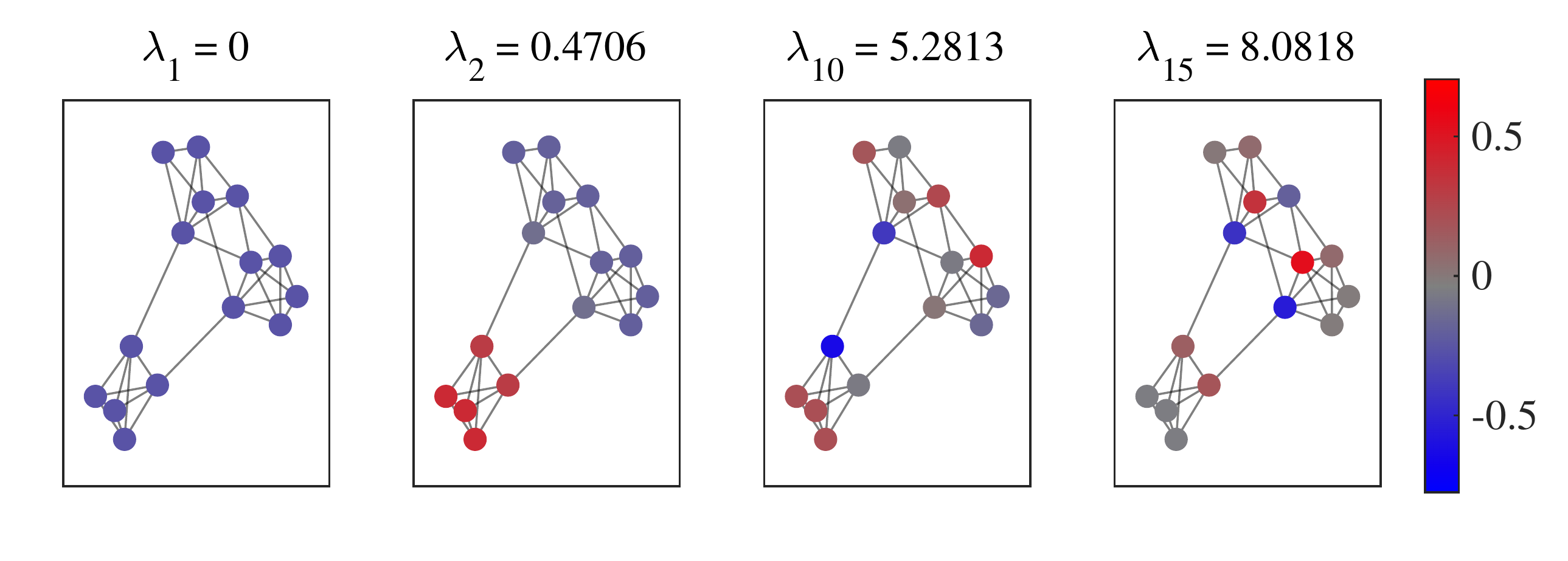}\vspace{-.3cm}
    \caption{The GFT basis, ${\bf u}_1, {\bf u}_2, {\bf u}_{10}, {\bf u}_{15}$, associated to the graph Laplacian of an undirected, unweighted graph with $15$ nodes. As the eigenvalue increases, the eigenvectors tend to be more oscillatory.} \vspace{-.2cm}  
    \label{fig:lowpass}
\end{figure*}

Let us take a closer look at the graph quadratic form ${\rm S}_2({\bm x})$. We set the eigendecomposition of the Laplacian matrix as ${\bm L} = {\bm U} \bm{\Lambda} {\bm U}^\top$ and assume that it has eigenvalues of multiplicity one ordered as $\bm{\Lambda} = \texttt{Diag}( \lambda_1 , \ldots, \lambda_n )$ with $0=\lambda_1 < \lambda_2 < \cdots < \lambda_n$, and 
${\bm U} = ( {\bf u}_1~{\bf u}_2~\cdots~{\bf u}_n )$ with ${\bf u}_i \in \mathbb{R}^{n}$ being the eigenvector for $\lambda_i$. 
Observe that for any ${\bm x} \in \RR^n$, it holds $\frac{{\rm S}_{2}( {\bm x} )}{\|\bm x\|_2} \geq 
\frac{ {\rm S}_{2}( {\bf u}_1 ) }{ \| {\bf u}_1 \|_2 } =\lambda_1$, and for any $\bm x$ orthogonal to  ${\bf u}_1$, it holds 
$\frac{{\rm S}_{2}( {\bm x} )}{\|\bm x\|_2} \geq \frac{{\rm S}_{2}( {\bf u}_2 )}{ \| {\bf u}_2 \|_2 } = \lambda_2$
and so on for the other eigenvectors. 
The observation indicates that the larger the eigenvalue, the more oscillatory the eigenvector is over the vertex set.
In particular, the smallest eigenvalue $\lambda_1=0$ is associated with the flat,  all-ones eigenvector ${\bf u}_1 = ( 1/\sqrt{n}) {\bf 1} $, as seen in Fig.~\ref{fig:lowpass}. 
The above motivates the definition of graph frequencies as the eigenvalues $\lambda_1,\ldots,\lambda_n$ and of the Graph Fourier Transform (GFT) basis  as the set of eigenvectors $\bm{U}$ \cite{sandryhaila2013discrete}. Therefore, the $i^{\text th}$ frequency component of ${\bm x}$ is defined as the inner product between ${\bf u}_i$ and ${\bm x}$:
\begin{equation}
\wt{x}_i = {\bf u}_i^\top {\bm x},~i=1,\ldots,n,
\end{equation}
and $\wt{\bm x} = {\bm U}^\top {\bm x}$ is called the GFT of ${\bm x}$. The magnitude of the GFT vector $|\wt{\bm x}|$ is the `spectrum' of the graph signal ${\bm x}$, where $| \wt{x}_i |^{2}$ represents the signal power at the $\lambda_i$th frequency. 

An important concept to modelling data with GSP is the graph filtering operation. To this end,
a linear \emph{graph filter} is described as the linear operator:
\beq \label{eq:graphfilter}
{\cal H}( {\bm L} ) = \sum_{ p=0}^{P-1} h_p {\bm L}^p = {\bm U} \Big( \sum_{p=0}^{P-1} h_p \bm{\Lambda}^p \Big) {\bm U}^\top,
\eeq
where $P$ is the \emph{filter order} (can be infinite) and $\{ h_p \}_{p=0}^{P-1}$ are the \emph{filter's coefficients}; as a convention, we use ${\bm L}^0 = \bm{I}$ and $ \lambda_{0}^{0} = 0^{0} = 1$. 
From \eqref{eq:graphfilter}, one can see that the graph filter has similar interpretation as an LTI filter in discrete-time signal processing where the former replaces the time shifts by powers of the GSO. Meanwhile, the second expression in \eqref{eq:graphfilter} defines the \emph{frequency response} as the diagonal matrix $h(\bm{\Lambda}) := \sum_{p=0}^{P-1} h_p \bm{\Lambda}^p$. 
A graph signal ${\bm y}$ is said to be filtered by ${\cal H}( {\bm L} )$ with the input excitation ${\bm x}$ when 
\begin{equation} \label{eq:gf_simple}
{\bm y} = {\cal H}({\bm L} ) {\bm x}.
\end{equation}
To better appreciate the effects of the graph filter, note that the $i$th frequency component of  ${\bm y}$ is:
\beq \label{eq:freq_rep}
\wt{y}_i = h( \lambda_i ) \cdot \wt{x}_i,~i=1, \ldots ,n,
\eeq 
where $h( {\lambda} ) := \sum_{p=0}^{P-1} h_p \lambda^p$ is the transfer function of the graph filter, or equivalently, we have $\wt{\bm y} = h( \bm{\lambda} ) \odot \wt{\bm x}$. 
It is similar to the convolution theorem in discrete-time signal processing.

\paragraph{Low-pass Graph Filter and Signal}
Inspired by \eqref{eq:freq_rep}, we define the ideal low-pass graph filter with a cut-off frequency $\lambda_k$ through setting the transfer function as $h(\lambda) = 1, \lambda \leq \lambda_k $ and $0$ otherwise \cite{tremblay2018design}.
Alternatively, one can say that a graph filter is low-pass if its frequency response is concentrated on the low graph frequencies. In this article, we adopt the following definition from \cite{wai2019blind}:\vspace{.1cm}

\fbox{\begin{minipage}{.99\linewidth}{\begin{Def} \label{def:lpf}
For any $1 \leq k \leq n-1$, define the ratio
\beq \label{eq:lpf}
\eta_k := \frac{ \max\{ | h( \lambda_{k+1} ) |, \ldots, |h( \lambda_n)| \} }{ \min\{ | h( \lambda_1 )|, \ldots, | h(\lambda_k) | \}  }.
\eeq
The graph filter ${\cal H}({\bm L})$ is $k$-low-pass if and only if the low-pass ratio $\eta_k$ satisfies $\eta_k \in [0,1)$.
\end{Def}}\end{minipage}}\vspace{.1cm}

The integer parameter $k$ characterizes the \emph{bandwidth}, or the cut-off frequency of the low-pass filter is at $\lambda_k$. The ratio $\eta_k$ quantifies the `strength' of the low-pass graph filter. 
Upon passing a graph signal through ${\cal H}({\bm L})$, the high frequency components (above $\lambda_k$) are attenuated by a factor of less than or equal to $\eta_k$. Using this definition, the ideal $k$-low-pass graph filter has the ratio $\eta_k = 0$, whose filter order has to be at least $P \geq n-k+1$ and transfer function  $h(\lambda)$
has  $\{\lambda_{k+1},\ldots,\lambda_{n}\}$ as its roots. 

Finally, a \emph{$k$-low-pass graph signal} refers to a graph signal that is the output of a $k$-low-pass filter, subject to a `well-behaved' excitation (i.e. does not possess strong high frequency components), which includes, but is not limited to, the white noise. 

\paragraph{The Impact of Graph Topologies} 
From Definition~\ref{def:lpf}, one can observe that the low-pass ratio $\eta_k$ of a graph filter depends on  the filter's coefficients $\{h_p \}_{p=0}^{P-1}$ and the graph Laplacian matrix's spectrum $\lambda_1, \ldots, \lambda_n$. The condition $\lambda_k \ll \lambda_{k+1}$ facilitates the design of a $k$-low-pass graph filter with a favorable ratio $\eta_k \ll 1$ and a low filter order $P$. As an example, the order-1 graph filter ${\cal H}({\bm L}) = {\bm I} - \lambda_n^{-1} {\bm L}$ is $k$-low-pass with the ratio $\eta_k = \frac{\lambda_n - \lambda_{k+1}}{\lambda_n - \lambda_k} = 1 - \frac{ \lambda_k - \lambda_{k+1} }{ \lambda_n - \lambda_k }$, where $\eta_k$ is small if $\lambda_k \ll \lambda_{k+1}$.

An example of graph topologies favoring the condition $\lambda_k \ll \lambda_{k+1}$ is the stochastic block model (SBM) \cite{rohe2011spectral} for describing random graphs with $k$ blocks/communities with nodes in  ${\cal N}$ partitioned as ${\cal N}_1,\ldots,{\cal N}_k$.
Consider a simplified SBM with $k$ equal-sized blocks specified by a membership matrix ${\bm Z} \in \{0,1\}^{n \times k}$ such that $Z_{i\ell} = 1$ if and only if $i \in {\cal N}_\ell$; and a latent model ${\bm B} \in [0,1]^{k \times k}$ where $B_{j,\ell}$ is the probability of edges between nodes in block $j$ and $\ell$. We consider the homogeneous planted partition model (PPM) such that ${\bm B} = b {\bf 1} {\bf 1}^\top + a {\bm I}$ with $b,a>0$. With the above specification, the adjacency matrix ${\bm A}$ is a symmetric binary matrix with independent entries satisfying $\EE[ {\bm A} ] = {\bm Z} \bm{B} {\bm Z}^\top$. 
When the graph size grows to infinity ($n \rightarrow \infty$), the Laplacian matrix of an SBM-PPM graph converges almost surely to its expected value \cite[Theorem 2.1]{rohe2011spectral}:
\beq
{\bm L} \overset{a.s.}{\longrightarrow} \EE[ {\bm L} ] = \frac{n (a + k b) }{k} \, {\bm I} - {\bm Z} \big( b{\bf 1} {\bf 1}^\top + a {\bm I} \big) {\bm Z}^\top.
\eeq
From the above, it can be shown that $\lambda_{k+1} - \lambda_k = \frac{na}{k} \gg 1$ for $\EE[ {\bm L} ]$, i.e., a favorable graph model for $k$-low-pass graph filters.
Lastly, the bottom-$k$ eigenvectors of the expected Laplacian associated with $\lambda_1, \ldots, \lambda_k$ can be collected into the matrix $\sqrt{ \frac{k}{n} } {\bm Z} {\bm P}$,
where ${\bm P}$ diagonalizes the matrix ${\bm B}$. In other words, the  eigenvectors corresponding to the bottom-$k$ eigenvalues of ${\bm L}$ will reveal the block structure. 

In contrast, the Erd\"{o}s-R\'{e}nyi graphs have Laplacian matrices that do not generally satisfy $\lambda_k \ll \lambda_{k+1}$. In fact, asymptotically ($n \rightarrow \infty$) the empirical distribution of the eigenvalues of Laplacian matrices tends to the free convolution of the standard Gaussian distribution and the Wigner’s semi-circular law \cite{ding2010spectral}. Such spectrum  does not favor the design of a $k$-low-pass graph filter with $\eta_k \ll 1$, reflecting the fact that block structure or communities do not emerge in Erd\"{o}s R\'{e}nyi graphs.

\paragraph{Low-pass Graph-Temporal Filter} 
When the excitation to a graph filter is of time-varying nature, and the topology is fixed, we consider 
a graph-temporal filter  \cite{isufi2016separable} with the impulse response:
\begin{align} 
 \textstyle {\cal H}(\bm L , t):= \sum_{p=0}^{P-1} h_{p,t}{\bm L}^{p} , \label{eq:timezgraph}
\end{align}
such that the graph filter's output is given by the time-domain convolution ${\bm y}_t = \sum_{s=0}^t {\cal H}( {\bm L}, t-s ) {\bm x}_s$. The filter is causal and $\bm{x}_s =\bm{0} $ for $s < 0$. We can apply $z$-transform and the GFT to the graph signal process $\{ {\bm x}_t \}_{t \geq 0}$ to obtain the $z$-GFT signal, $\wt{\bm X}(z)$, given by:
\begin{align}
 \textstyle   \bm{X}(z) = \sum_{t=0}^{\infty} {\bm x}_t z^{-t}, ~~~ \wt{\bm X}(z) = {\bm U}^\top {\bm X}(z), \label{eq:z_GFT}
\end{align}
which represents $\{ {\bm x}_t \}_{t \geq 0}$ in the joint $z$-graph frequency domain. 
With that, we obtain the input-output relation $\wt{\bm{Y}}(z)= \wt{\bm h}(z) \odot \wt{\bm{X}}(z)$ and graph-temporal joint transfer function $\mathbb{H}(\lambda,z) := \sum_{t=0}^{\infty}\sum_{p=0}^{P-1}h_{p,t}\lambda^p z^{-t}$.
A  class of graph-temporal filters for modeling graph signal processes is the GF-ARMA $(q,r)$ filter, whose input-output relation in time domain and $z$-GFT domain are described below, respectively:
\begin{equation} \notag
\begin{split}
& {\bm y}_t - {\cal A}_1( {\bm L} ) {\bm y}_{t-1}  \cdots - {\cal A}_q ({\bm L}) {\bm y}_{t-q} \\
& \hspace{.5cm} = {\cal B}_0( {\bm L} ) {\bm x}_t + \cdots + {\cal B}_r( {\bm L} ) {\bm x}_{t-r}, \\
&
\wt{\bm a}(z) \odot \wt{\bm{Y}}(z)  = \wt{\bm b}(z) \odot \wt{\bm{X}}(z) ,
\end{split}
\end{equation}
where $\wt{\bm a}(z)=1-\sum_{s=1}^q \wt{\bm a}_s z^{-s}$ and  $\wt{\bm b}(z)=\sum_{s=0}^r \wt{\bm b}_s z^{-s}$ are the $z$-transform of the graph frequency responses of the graph filter taps $\{{\cal A}_s ({\bm L}) \}_{s=1}^q$, $\{{\cal B}_s ({\bm L}) \}_{s=0}^r$ for the GF-ARMA $(q,r)$ filter. 
Note that the  joint frequency response is given by $ \mathbb{H}( \lambda_i, z ) = {[\wt{\bm b}(z)]_{i}} \big/ {[\wt{\bm a}(z)]_{i}}$, whose poles and zeros may vary depending on the graph frequencies $\lambda_1, \ldots, \lambda_n$. 
A relevant case is when ${\cal H}(\bm L ,t)$ is a \emph{low-pass graph-temporal filter}. Similar to Definition~\ref{def:lpf}, we say that ${\cal H}(\bm L ,t)$ is low-pass with a cutoff frequency $(\lambda_k, \omega_0)$ and ratio $\eta_k$ if:
\beq \label{eq:lpgtf}
\eta_k = \frac{ \max_{ \lambda \in \{ \lambda_{k+1}, \ldots, \lambda_n \}, \omega \in (\omega_0, 2\pi) } |\mathbb{H}(\lambda, e^{j \omega})| }{ \min_{ \lambda \in \{ \lambda_{1}, \ldots, \lambda_k \}, \omega \in [0, \omega_0] } |\mathbb{H}(\lambda, e^{j \omega})| } < 1. 
\eeq 
Graph signals filtered by a low-pass graph-temporal filter are also commonly found in applications, as we will illustrate next.

\vspace{-0.3cm}

\section{Models of Low-pass Graph Signals}\label{sec:model}
Before studying the GSP tools for low-pass graph signals, a natural question is where can one find such graph signals? It turns out that many physical and social processes are naturally characterized by low-pass graph filters.
In this section, we present various examples and show that their generation processes can be represented as outputs from low-pass graph filters.

\paragraph{Diffusion Model} The first case pertains to   observations from a diffusion process, whose variants are broadly applicable in network science. As an example, we consider the heat diffusion model in \cite{thanou2017learning}. In this example, the relevant graph is a proximity graph where each node $i \in {\cal N}$ is a location (e.g., cities), and if locations $i,j$ are close to each other, then $(i,j) \in {\cal E}$. The graph is endowed with a symmetric weighted adjacency matrix encoding the distance between locations. The graph signal ${\bm y}_t \in \RR^n$ encodes the temperature of $n$ locations at time $t$, and let ${\bm x}_0 \in \RR^n$ be the initial heat distribution. The temperature of a location is diffused to its neighbors. Let $\sigma > 0$ be a constant,  we have 
\vspace{-.1em}
\beq \label{eq:diff_model}
{\bm y}_{t} = e^{-t \sigma {\bm L}} {\bm x}_0 = \big( {\bm I} - t \sigma {\bm L} + \frac{(t \sigma)^2}{2} {\bm L}^2 - \cdots \big) {\bm x}_0
\eeq
where \eqref{eq:diff_model} is a discretization of the heat diffusion equation \cite{thanou2017learning}. 
As ${\bm L} {\bf 1} = {\bf 0}$, the matrix exponential $e^{-t \sigma {\bm L}} = {\bm I} - t \sigma {\bm L} + \frac{(t \sigma)^2}{2} {\bm L}^2 - \cdots$ is row stochastic.  
The temperature at time $t$ is thus 
a weighted average of neighboring locations' temperatures at  $t=0$, i.e., this is a diffusion dynamical process. 

To understand \eqref{eq:diff_model} under the context of low-pass filtering, we observe that ${\bm y}_{t}$ is a filtered graph signal with the excitation ${\bm x}_0$ and the graph filter ${\cal H}({\bm L}) = e^{-t \sigma {\bm L}}$.
We verify that ${\cal H}({\bm L})$ is $k$-low-pass with Definition~\ref{def:lpf} for any $k < n$. Note that the low-pass ratio $\eta_k$ is:
\beq \notag
\frac{ e^{- t \sigma \lambda_{k +1}} }{ e^{- t \sigma \lambda_{k }} } = e^{-t \sigma( \lambda_{k+1} - \lambda_k) } . 
\eeq
As $\lambda_{k +1} > \lambda_{k }$ and $t \sigma >0$, we see that ${\cal H}({\bm L})$ is a $k$-low-pass graph filter for any $k=1,\ldots,n-1$.

We have assumed that ${\bm x}_0$ is an impulse excitation affecting the system only at the initial time. In practice, the excitation signal may not be an impulse and the output graph signal ${\bm y}_t$ is expressed as the convolution ${\bm y}_t = \sum_{s=0}^{t} e^{-(t-s)\sigma {\bm L}} {\bm x}_s$. This corresponds to a low-pass graph-temporal filter  with the joint transfer function $\mathbb{H}(\lambda,z) = (1 - e^{-\lambda \sigma} z^{-1})^{-1}$.
Besides, the diffusion process is common in network science as similar models arise in contagion process and product adoption to name a few.

\paragraph{Opinion Dynamics} \label{case:op}This example pertains to opinion data mined from social networks with the influence of external excitation \cite{wai2019blind,ravazzi2017learning}. 
The relevant graph ${\cal G}$ is the social network graph where each node $i \in {\cal N}$ is an individual, and ${\cal E}$ is the set of friendships. Similar to the previous case study, this graph is endowed with a symmetric weighted adjacency matrix ${\bm A}$, where the weights measure the trust among pairs of individuals. Let $\alpha \in (0,\lambda_n^{-1})$, $\beta \in (0,1)$ be parameters of trust on others and susceptibility to external influence of an individual respectively. The evolution of opinions follows that of a combination of DeGroot's and Friedkin-Johnsen's model \cite{friedkin2011formal}, which is a GF-AR(1) model: 
\beq \label{eq:op_dyn}
{\bm y}_{t+1} = (1-\beta) \big( {\bm I} - \alpha {\bm L} \big) {\bm y}_t + \beta {\bm x}_t ,
\eeq
where ${\bm y}_t \in \RR^n$ is a graph signal of the individuals' opinions at time $t$, and ${\bm x}_t \in \RR^n$ is a graph signal of the external opinions perceived by the social network. Note that this also corresponds to a low-pass graph-temporal filter  with the joint transfer function $\mathbb{H}(\lambda,z) = \beta [1 - (1-\beta)(1-\alpha \lambda) z^{-1}]^{-1}$.

To discuss the steady state of \eqref{eq:op_dyn}, let us assume that ${\bm x}_t \equiv \bm{x}$. Considering \eqref{eq:op_dyn}, we observe that ${\bm y}_{t+1}$ is  a convex combination of ${\bm x}$ and weighted average of the neighbors' opinions at time $t$ that is formed by taking a weighted average of neighboring signals in ${\bm y}_t$ using a diffusion operator ${\bm I} - \alpha {\bm L}$. As $\beta > 0$, the recursion is stable, leading to the steady state (or equilibrium) opinions:
\beq \label{eq:steady}
{\bm y} = \lim_{t \rightarrow \infty} {\bm y}_t = \left( {\bm I} +  \wt \alpha  {\bm L} \right)^{-1} {\bm x} = {\cal H}({\bm L}) {\bm x},
\eeq
where we have defined $\wt\alpha =  \beta (1-\alpha) / \alpha > 0$ and ${\bm y}$ is a filtered graph signal excited by ${\bm x}$.

The graph filter  above is given by ${\cal H}({\bm L}) = ( \bm{I} + \wt\alpha   {\bm L} )^{-1}$.
To verify that it is a $k$-low-pass graph filter with Definition~\ref{def:lpf}, we note that for any $k < n$, the low-pass ratio $\eta_k$ is
\beq \notag
\frac{ 1 + \wt\alpha \lambda_{k} }{ 1 + \wt\alpha \lambda_{k+1} } = 1 - \wt\alpha \frac{\lambda_{k+1}-\lambda_k}{1 + \wt\alpha \lambda_k }.
\eeq
Again, we observe that as $\lambda_{k+1} > \lambda_k$, the above graph filter is $k$-low-pass for any $k=1,\ldots,n-1$. However, we remark that this low-pass ratio may be undesirable with $\eta_k \approx 1$ when $\wt\alpha \ll 1$. Interestingly, a similar generative model as \eqref{eq:op_dyn} is found in equilibrium problems such as quadratic games \cite{candogan2012optimal}. 

Two remarks are in order. First, social networks are typically directed, and this suggests using a non-symmetric shift operator as opposed to the symmetric Laplacian matrix, which we used for simplicity of exposition. Second, many alternative models for social networks interactions are non-linear and linear GSP is insufficient in those contexts.

\paragraph{Finance Data} Financial systems such as stock market and hedge funds produce return reports periodically about their business performances. A collection of these reports can be studied as graph signals, where the relevant graph ${\cal G}$ consists of nodes ${\cal N}$ that are financial institutions, and edges ${\cal E}$ that are business ties between them. It has been studied  \cite{billio2012econometric} that business performances are correlated according to the business ties. Moreover, the returns are affected by a number of common factors \cite{mantegna1999introduction}. Inspired by  \cite{billio2012econometric}, \cite[Ch.~12.2]{mantegna1999introduction}, let $\beta \in (0,1)$ be the strength of external influences, a reasonable model for the transient dynamics of the graph signal ${\bm y}_t$ of business performance measures is also a   GF-AR(1):
\beq \label{eq:fin_sys}
{\bm y}_{t+1} = (1-\beta) {\cal H}({\bm L}) {\bm y}_t + \beta {\bm B} {\bm x},
\eeq
where ${\cal H}({\bm L})$ is an unknown but low-pass graph filter, ${\bm B} \in \RR^{n \times r}$ represents the factor model affecting financial institutions, and ${\bm x} \in \RR^r$ is the excitation strength. The equilibrium of \eqref{eq:fin_sys} is:
\beq \notag
{\bm y} = \lim_{t \rightarrow \infty} {\bm y}_t = \Big( \frac{1}{\beta} {\bm I} - \frac{\ol\beta}{\beta} {\cal H}( {\bm L} ) \Big)^{-1} {\bm B} {\bm x} \equiv \wt{\cal H}( {\bm L} ) {\bm B} {\bm x},
\eeq
where $\ol\beta = 1-\beta$. We see that ${\bm B}{\bm x}$ is the excitation signal and the equilibrium ${\bm y}$ is the filter output. 
Suppose that ${\cal H}( {\bm L} )$ is a $k$-low-pass graph filter with the frequency response satisfying $h(\lambda) \geq 0$, then for $\wt{\cal H}({\bm L})$, we can evaluate the low-pass ratio $\eta_k$ as
\beq \notag
\begin{split}
1 - \frac{\ol\beta \big\{ \min_{ \ell=1,\ldots,k } h( \lambda_\ell) - \max_{ \ell=k+1,\ldots,n } h( \lambda_\ell) \big\} }{1 - \ol\beta  \max_{ \ell=k+1,\ldots,n } h( \lambda_\ell) } .
\end{split}
\eeq
As $\min_{ \ell=1,\ldots,k } h( \lambda_\ell) - \max_{ \ell=k+1,\ldots,n } h( \lambda_\ell) > 0$ since 
${\cal H}({\bm L})$ is a $k$-low-pass graph filter itself, we observe that $\wt{\cal H}( {\bm L} )$ is again $k$-low-pass according to Definition~\ref{def:lpf}.

For ${\bm y}$ to be a $k$-low-pass graph signal, one has to also assume that ${\bm B} {\bm x}$ is not high-pass (i.e., not orthogonal to a low-pass one). This is a mild assumption as the latent factor affecting financial institutions are either independent of the network, or are aligned with the communities. Above all, we remark that \eqref{eq:fin_sys} is an idealized model where determining the exact model is an open problem in economics, see   \cite{mantegna1999introduction,billio2012econometric}.

\paragraph{Power Systems} \label{para:power}
In the case of power systems, the relevant graph ${\cal G} = ({\cal N},{\cal E}) $  is the electrical transmission  lines network. The node (a.k.a.~bus) set includes generator buses, ${\cal N}_g = \{1,\ldots,|{\cal N}_g| \}$, and non-generator/load buses, ${\cal N}_{\ell} = {\cal N} \setminus {\cal N}_g$. The edge set $\mathcal{E}$ refers to the transmission lines connecting the buses. 
The \textit{branch admittance matrix}, $\bm{Y}$,  models  the effect of transmission lines and is a complex symmetric matrix associated with ${\cal G}$, where $\left[\bm{Y}\right]_{ij}$ is the complex admittance of the branch between nodes $i$ and $j$ provided that  $(i,j) \in \mathcal{E}$    .   
The graph signals we consider are the complex voltage phasors, denoted as ${\bm v}_t \in \CC^n$ when measured at time $t$. They can be obtained using phasor measurement units (PMU) \cite{glover2008power} installed on each bus $i \in {\cal N}$. The graph shift operator in this case is a diagonally perturbed branch admittance matrix:
\begin{align}
{\bm S}_{\sf grid} :=  \bm{Y} + \texttt{diag} \left( {\small [
\bm{y}_{g}^{\top}, 
\bm{y}_{\ell}^{\top}(0)] } \right),
\label{eq:grid_GSO}
\end{align}
where $\bm{y}_{g} \in \CC^{|{\cal N}_g|}$ is the generator admittance and ${\bm y}_\ell(0) \in \CC^{|{\cal N}_\ell|}$ is the load admittance at $t=0$. 

Note that ${\bm S}_{\sf grid}$ is a GSO on the grid graph as $[{\bm S}_{\sf grid}]_{ij} = 0$ if $(i,j) \notin {\cal E}$. The \emph{complex symmetric} matrix ${\bm S}_{\sf grid}$ can be decomposed as ${\bm S}_{\sf grid} = {\bf U} \bm{\Lambda} {\bf U}^\top$, where ${\bf U}$ is a complex orthogonal matrix satisfying ${\bf U}^\top {\bf U} = {\bm I}$, and $\bm{\Lambda}$ is a diagonal matrix with diagonal elements $\{ \lambda_1, \ldots, \lambda_n \}$ sorted as $0 < | \lambda_1 | \leq \cdots \leq |\lambda_n|$; see \cite{ramakrishna2019modeling} for modeling details.
Let ${\bm i}_t^g \in \CC^n$ be the outgoing current at each node at time $t$, given by $\bm{i}^{g}_{t} :=
  [\bm{y}^\top_{g} \odot {\exp\left(\bm{x}^\top_{t}\right)}, \bm{0}]^\top,$
where elements in $\exp\left(\bm{x}_{t}\right) \in \mathbb{C}^{\abs{\mathcal{N}_g}}$ are the internal voltage phasors  at the generator buses.  
Applying Kirchoff's current law in quasi-steady state, the voltage phasors $\bm{v}_{t} \in \CC^n$ are:
\begin{align}
\bm{v}_{t} = {\bm S}_{\sf grid}^{-1} \, \bm{i}^{g}_{t} + \bm{w}_{t} = \mathcal{H}( {\bm S}_{\sf grid}) \, \bm{i}^{g}_{t} + \bm{w}_{t}, \label{eq:generative_model_new_GSO}
\end{align} 
where ${\bm w}_t \in \CC^n$ captures the slow time-varying nature of the load and other modeling approximations. In other words, ${\bm v}_t$ is a graph signal obtained by the graph filter ${\cal H}({\bm S}_{\sf grid}) = {\bm S}_{\sf grid}^{-1}$ and the excitation signal ${\bm i}_t^g \in \CC^n$. 
Particularly, we observe that ${\cal H}({\bm S}_{\sf grid}) = {\bm S}_{\sf grid}^{-1}$ is a low-pass graph filter. Consider any $k \leq n$, the low-pass ratio $\eta_k$ is
\begin{equation} \notag
  {|\lambda_{k}|} / {|\lambda_{k+1}|} .
\end{equation}
As the power grids tend to be organized as \textit{communities} to serve different areas with high population densities, the system admittance matrix $\bm{Y}$ is block diagonal and  sparse. In particular, with $k$ communities in the grid graph, these facts indicate that the graph filter is $k$-low-pass satisfying $\eta_k \ll 1$. 

The excitation graph signal $\bm{i}^{g}_{t}$ itself has a low-rank structure, as $[\bm{i}^{g}_{t}]_i = 0 $ at $i \notin {\cal N}_g$. The  temporal dynamics of $\bm{x}_{t}$  can be described as an AR(2) graph filter \cite{isufi2016separable} using a reduced \textit{generator-only} shift operator $ \bm{S}_{\text{red}} \in \mathbb{C}^{\abs{\mathcal{N}_g}}$ with the graph-temporal transfer function, $\mathbb{H}(\lambda_{\text{red}},z)$, 
\beq \notag
\begin{split}
& \bm{x}_{t} = \sum_{s=0}^t \mathcal{H} \left( \bm{S}_{\text{red}}, t-s \right) \bm{p}_{s},\\
& \mathbb{H}(\lambda_{\text{red}},z) = \sigma_{p}^{2} \Big(1- \sum_{p=0}^{1} a_{p,1}\lambda_{\text{red}}^{p} z^{-1}-\sum_{p=0}^{1} a_{p,2}\lambda_{\text{red}}^{p} z^{-2}\Big)^{-1}. 
\end{split}
\eeq
where $\bm{p}_{s}$ is the stochastic power input to the system. The graph-temporal filter is low-pass in the time domain. The overall system in \eqref{eq:generative_model_new_GSO} has approximately  the properties of a low-pass graph temporal filter according to the definition in \eqref{eq:lpgtf}. 

\section{User Guide to Low-pass Graph Signal Processing} \label{sec:tools}

If we observe a set of low-pass graph signals such as those from Section \ref{sec:model}, what can we learn from these signals? Can we find efficient representations for them? Can we exploit this structure to denoise the signal or detect anomalies? 
To answer these questions, we begin by studying two salient features of low-pass graph signals, namely \emph{low-rank covariance matrix}, and \emph{smoothness} as measured by the graph quadratic form. Then, we illustrate how these features can enable low-pass GSP to sample graph signals (and therefore compress them), to infer the graph topology, and to detect anomalous activities. Furthermore, when the graph topology admits a clustered structure, we highlight how these clusters emerge in the low-pass graph signals and provide insights on the optimal sampling patterns. 

We now consider a set of $m$ low-pass graph signals that can be modeled as outcomes of independent and identically distributed {\it random experiments}, given as:
\beq \label{eq:lpf_graphsignal}
{\bm y}_\ell = {\cal H}({\bm L}) \, {\bm x}_\ell + {\bm w}_\ell, \quad \ell =1,\ldots,m,
\eeq
such that ${\cal H}({\bm L})$ is a $k$-low-pass graph filter defined on the Laplacian matrix, ${\bm x}_\ell$ is the excitation signal, and ${\bm w}_\ell$ is an additive noise. 
For simplicity, we do not consider the more general low-pass graph-temporal processes and assume that ${\bm x}_\ell$, ${\bm w}_\ell$ are zero-mean white noise with $\EE[ {\bm x}_\ell {\bm x}_\ell^\top ] = {\bm I}$, $\EE[ {\bm w}_\ell {\bm w}_\ell^\top ] = \sigma^2 {\bm I}$. We remark that the following observations still hold for the general setting when $\EE[ {\bm x}_\ell {\bm x}_\ell^\top ]$ is not white, or even diagonal. The latter relaxation is important for the applications listed in Section~\ref{sec:model}. For instance, in opinion dynamics, the excitation signals may represent external opinions that do not affect the social network uniformly, e.g., they are news articles written in a foreign language.

\subsection{{Low-rank Covariance Matrix}}
From \eqref{eq:lpf_graphsignal}, it is straightforward to show that $\{ {\bm y}_\ell \}_{\ell=1}^m$ is zero-mean with the covariance matrix
\beq \label{eq:lowrank_cov}
{\bm C}_y = {\bm U} h( \bm{\Lambda} )^2 {\bm U}^\top + \sigma^2 {\bm I} .
\eeq 
Recall that ${\cal H}({\bm L})$ is a $k$-low-pass graph filter, if $\eta_k \ll 1$ as defined in \eqref{eq:lpf}, the energy of $h( \bm{\Lambda} )$ will be concentrated in the top-$k$ diagonal elements.
Therefore, when the noise variance is small ($\sigma^2 \approx 0$), the low-pass graph signals lie approximately in $\mbox{span}({\bm U}_k)$, a $k$-dimensional subspace of $\RR^n$. 

\paragraph*{a-i). Sampling Graph Signals}
As the $k$-low-pass graph signals lie approximately in $\mbox{span}({\bm U}_k)$, it is possible to map the graph signals almost losslessly onto  $k$-dimensional vectors. While the 
$k$-dimensional representation can be obtained by projecting on the space spanned by ${\bm U}_k$, it is not necessary to do so. 
An alternative to generate this $k$-dimensional representation is by decimating the graph signals. 

\begin{figure*}
    \centering \begin{tabular}{r c r c r c}
    \addvbuffer[0cm 4.75cm]{\sf (a)}\hspace{-.45cm}
    & \includegraphics[width=0.29\textwidth]{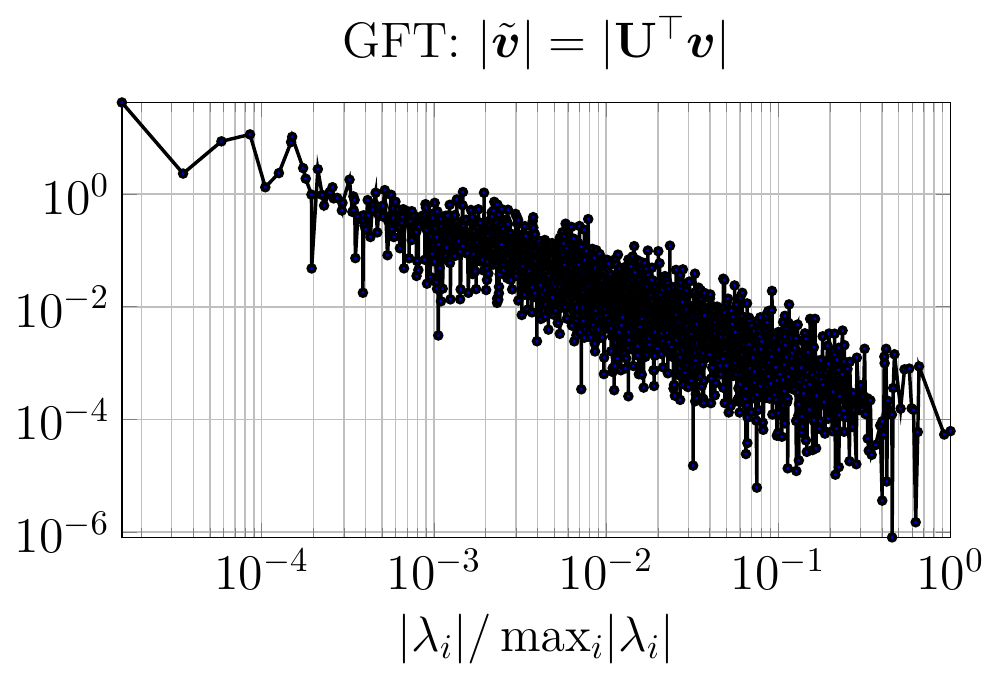}
    & \addvbuffer[0cm 4.75cm]{\sf (b)}\hspace{-.45cm}
    & \includegraphics[width=0.29\textwidth]{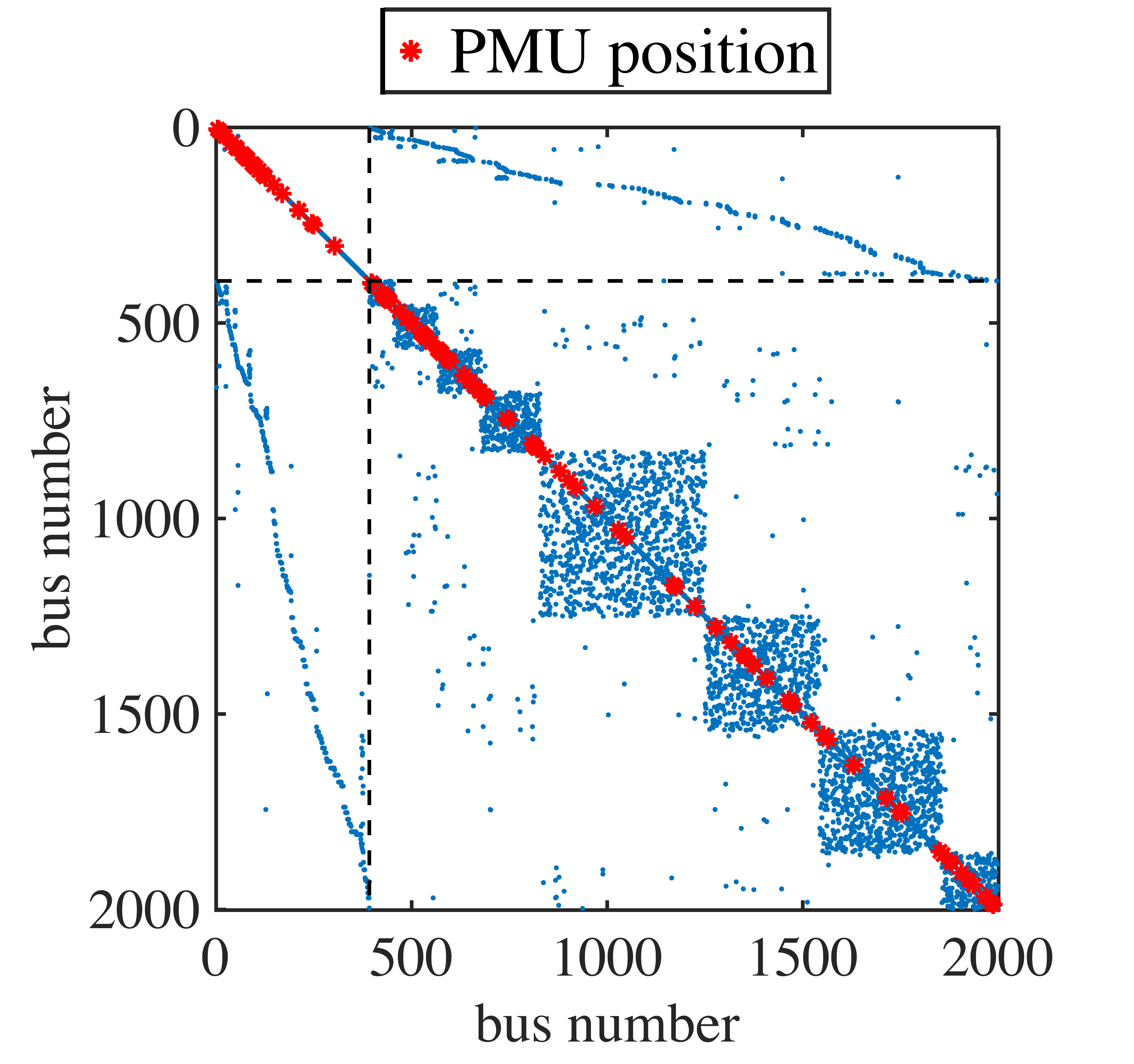}
    & \addvbuffer[0cm 4.75cm]{\sf (c)}\hspace{-.45cm}
    & \includegraphics[width=0.27\textwidth]{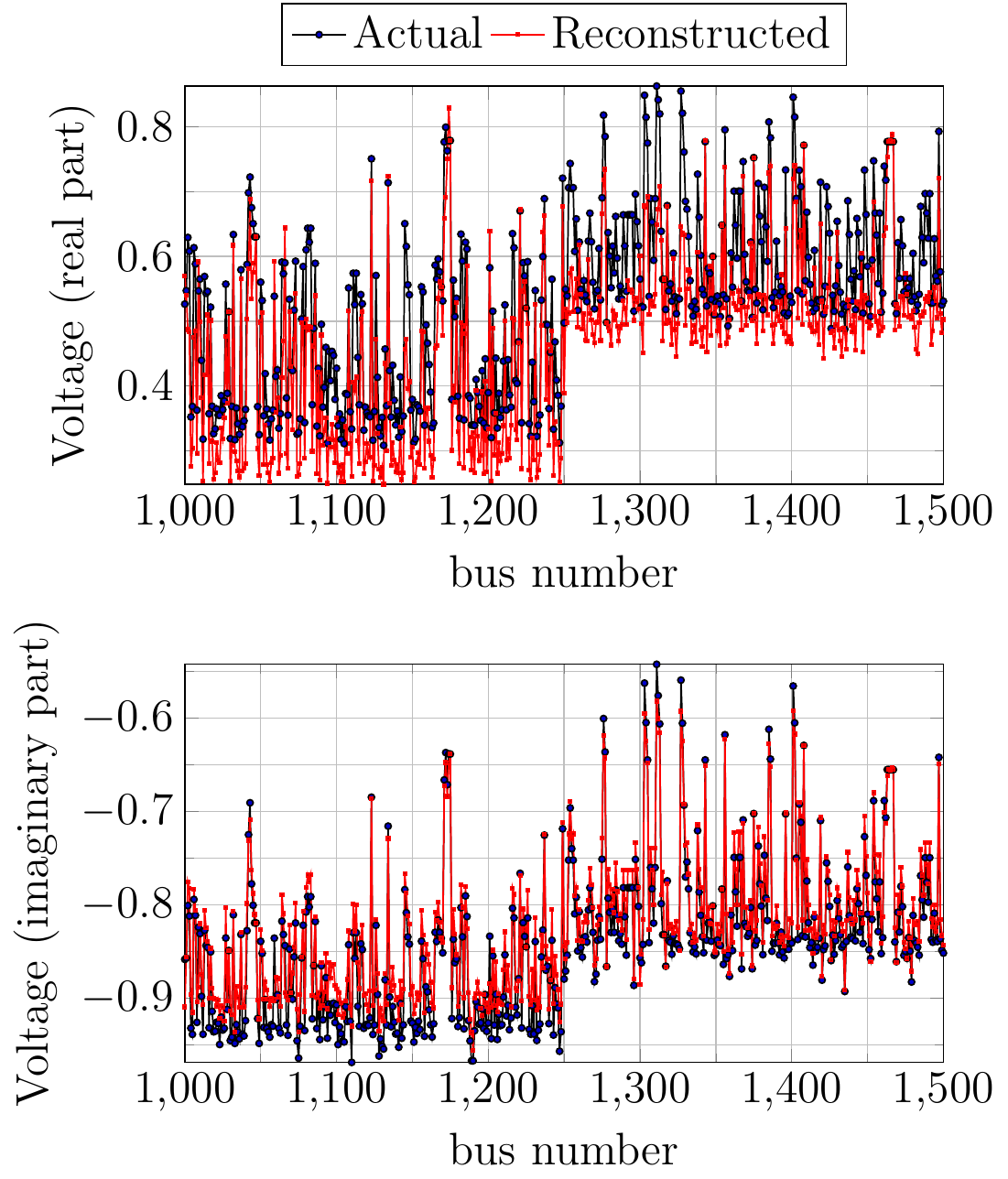}
    \end{tabular}
    \vspace{-.2cm}
    \caption{({\sf a}) Magnitude of Graph Fourier Transform (GFT) of voltage graph signal plotted with respect to normalized graph frequencies. ({\sf b}) Sampling pattern overlaid on the support of GSO $\bm{S}_{\sf grid}$  for placing synchro-phasors sensors on the synthetic 2000 bus ACTIVSg power grid system by employing a greedy method to find $k=100$ rows of $\mathbf{U}_{k}$ so that the smallest singular value of $\bm{\Phi}\bm{U}_{k} $ is maximized  (The first entries correspond to generator buses). This case emulates the grid for the state of Texas (called ERCOT systems), where there are 8 areas, matching the number of communities evident from the GSO. ({\sf c}) Reconstructed voltage graph signal using optimally placed sensors at a subset of buses (1000-1500). }\vspace{-.2cm}
    \label{fig:activs-sampling}
\end{figure*}

To describe the setup, we let ${\cal N}_s = \{ s_1, \ldots, s_{n_s} \} \subset {\cal N}$ be a \textit{sampling} set with cardinality $n_s = |{\cal N}_s|$. A sampled version of ${\bm y}_\ell$ is constructed as (omitting the subscript $\ell$ for brevity):
\beq \label{eq:sampling}
\text{{\sf (i)} select a subset ${\cal N}_s \subset {\cal N}$; {\sf (ii)} set}~~{\bm y}_{\sf samp} = \bm{\Phi} {\bm y},
\eeq 
such that
\[
[\bm{\Phi}]_{q,j} = \begin{cases}
1, & \text{if}~j = s_q,  \\
0, & \text{otherwise},
\end{cases}
\]
and $\bm{\Phi} \in \RR^{ n_s \times n}$ is a fat sampling matrix that compresses the graph signal to an $n_s$-dimensional vector. 
To recover ${\bm y}$, we interpolate ${\bm y}_{\sf samp}$ using a matching linear transformation \cite{chen2015discrete}, giving $\wh{\bm y} = \bm{\Psi} {\bm y}_{\sf samp}$ with $\bm{\Psi} \in \RR^{n \times n_s}$ to be designed later. Clearly, when $n_s < n$, it is not possible to exactly recover an \emph{arbitrary graph signal}. To ensure exact recovery, we see that it requires certain additional conditions on the sampling set and the graph signal.  

Exactly recovering ${\bm y}$ from its sampled version ${\bm y}_{\sf samp}$ would require the sampled graph signal to be in the range space of sampling matrix $\bm{\Phi}$.
We let $\ol{\bm y} = {\bm U}_k {\bm U}_k^\top {\bm y}$ be the projection of ${\bm y}$ onto the (low-frequency) subspace spanned by ${\bm U}_k$ and $\ol{\bm w} = {\bm y} - \ol{\bm y}$ be the projection error.
The projected graph signal $\ol{\bm y}$ is a $k$-bandlimited (in fact, $k$-low-pass) graph signal. From \cite[Theorem 1]{chen2015discrete}, a sufficient condition for exact recovery is that if
\beq \label{eq:sampling_cond}
{\rm rank} \big( \bm{\Phi} {\bm U}_k \big) = k,
\eeq
then there exists an interpolation matrix $\bm{\Psi} \in \RR^{n \times n_s}$ such that $\bm{\Psi} \bm{\Phi} \ol{\bm y} = \ol{\bm y}$. In fact, it is possible to recover any $k$-bandlimited graph signal from its sampled version. We have
\beq \label{eq:sampling_recons}
\wh{\bm y} = \bm{\Psi} {\bm y}_{\sf samp} = \ol{\bm y} + \bm{\Psi} \bm{\Phi} \ol{\bm w} = {\bm y} + (  \bm{\Psi} \bm{\Phi} - {\bm I} ) \ol{\bm w}.
\eeq 
As $k$-low-pass graph signals lie approximately in ${\rm span}({\bm U}_k)$, we have $\ol{\bm w} \approx 0$ provided that $\eta_k \ll 1$. Under condition \eqref{eq:sampling_cond}, the sampling-and-interpolation procedure results in a small interpolation error.

Clearly, a necessary condition to satisfy \eqref{eq:sampling_cond} is $n_s \geq k$, i.e., we require at least the same number of samples as the bandwidth of the low-pass graph filter which produces the graph signal ${\bm y}$. Beyond the necessary condition, obtaining a sufficient condition for \eqref{eq:sampling_cond} can be difficult as it is not obvious to derive conditions on the sampling set. The design of the sampling set has been the focus of work in \cite{chen2015discrete,anis2016efficient,tsitsvero2016signals} which propose to find ${\cal N}_s$ via a greedy method, or via the graph spectral proxies.
The above statements are valid for any graph signal that has a sparse frequency support. In the case of low-pass graph signals, we can obtain insights on what type of sampling patterns are compatible with \eqref{eq:sampling_cond}. Consider the special case of SBM-PPM graphs with $k$ blocks discussed in Section~\ref{sec:basic}. Note that as $n \rightarrow \infty$, we have ${\bm U}_k = \sqrt{k/n} \, {\bm Z} {\bm P}$ for this model, where ${\bm Z}$ is the block-membership matrix. Subsequently, condition  \eqref{eq:sampling_cond} can be easily verified if ${\cal N}_s$ contains \emph{at least one node} from \emph{each of the $k$ blocks}.

In Fig. \ref{fig:activs-sampling},  we consider a power system application. We first plot the magnitude of GFT of voltage graph signal with respect to normalized graph frequencies in log scale. From
the linear decay, it is evident that the magnitude of GFT coefficients at lower frequencies is higher confirming the signal is low-pass in nature. Then, the sampling pattern (or optimal placement of sensors) for graph signal reconstruction is shown in the figure.
The block  structure in the GSO for the electric grid guides the sampling strategy. In this example, the smallest singular value of $\bm{\Phi}\bm{U}_{k} $ is maximized using a greedy algorithm \cite{tsitsvero2016signals}.

\paragraph*{a-ii). Blind Community Detection}
Another consequence of \eqref{eq:lowrank_cov} relates to learning the \emph{block or community structure} when the graph topology is unknown. When the graph topology is known, spectral clustering (SC) is often the method of choice. The SC method computes the bottom-$k$ eigenvectors of Laplacian as ${\bm U}_k$ and partitions the $n$ nodes via $k$-means:
\beq \label{eq:kmeans}
F^\star = \min_{ {\cal N}_1,\ldots, {\cal N}_k }~ F({\cal N}_1,\ldots, {\cal N}_k ; {\bm U}_k ) ,
\eeq
where
{\small\beq \notag
F({\cal N}_1,\ldots, {\cal N}_k ; {\bm U}_k ) := \Big( \sum_{q=1}^k \sum_{i \in {\cal N}_q } \Big\| {\bf u}_i^{\rm row} - \frac{1}{|{\cal N}_q|} \sum_{j \in {\cal N}_q} {\bf u}_j^{\rm row} \Big\|_2^2 \Big)^{1/2},
\eeq}
such that ${\bf u}_i^{\rm row} \in \RR^k$ is the $i$th row vector of ${\bm U}_k$. 
In fact, this is an effective method for SBM-PPM graphs where solving \eqref{eq:kmeans} reveals the true block membership \cite{rohe2011spectral}.

\begin{figure*}[t]
    \centering\begin{tabular}{c c r c}
    \addvbuffer[0cm 5cm]{\sf (a)}\hspace{-.45cm} &\includegraphics[width=.225\linewidth]{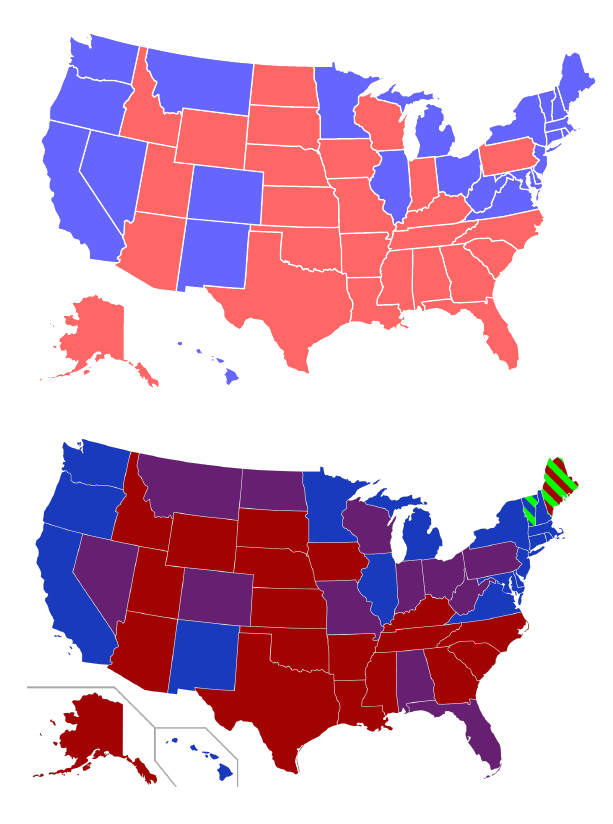}~& \addvbuffer[0cm 5cm]{\sf (b)}\hspace{-.95cm}&\hspace{-.5cm} \includegraphics[width=.475\linewidth]{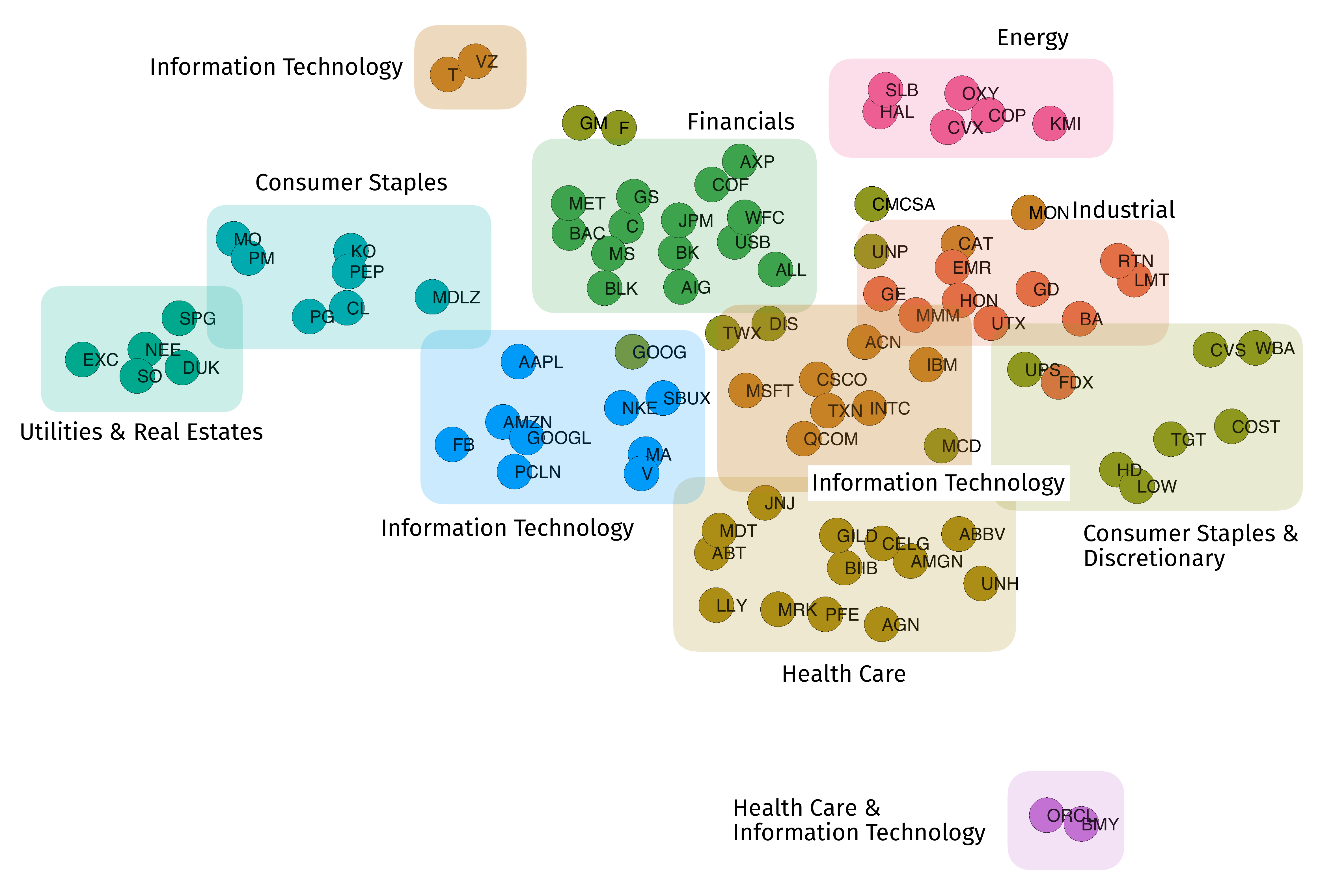}\end{tabular}\vspace{-.1cm}
    \caption{Communities detected from opinion dynamics and stock data. {\sf (a)} US Senate voting records: (Top) inferred membership via BlindCD, (Bottom) actual party membership of Senators taken from [\url{https://en.wikipedia.org/wiki/115th_United_States_Congress}]. Note that the purple color in the bottom indicates that the state has a Democrat Senator and a Republican Senator. {\sf (b)} Daily returns of S\&P100 stocks. Colors on the nodes represent different detected communities. Communities are manually labeled according to business types.}
    \label{fig:senate_stock}\vspace{-.2cm}
\end{figure*}

Although only the graph signals $\{ {\bm y}_\ell \}_{\ell=1}^m$ are observed, we know from \eqref{eq:lowrank_cov} that the covariance matrix ${\bm C}_y$ will be dominated by a rank-$k$ component spanned by ${\bm U}_k$ under the low-pass assumption. 
In fact, this is precisely what we need for community detection as hinted in \eqref{eq:kmeans}.
To this end, \cite{wai2019blind} proposed the blind community detection (BlindCD) procedure: 
\[
\begin{array}{l}
\text{{\sf (i)} find the top-$k$ eigenvectors $\wh{\bm U}_k \in \RR^{n \times k}$}\\[.1cm]
~~~~\text{of sample covariance $\wh{\bm C}_y = \frac{1}{m} \sum_{\ell=1}^m {\bm y}_\ell {\bm y}_\ell^\top$;} \\[.1cm]
\text{{\sf (ii)} apply $k$-means on the rows of $\wh{\bm U}_k$.}
\end{array}
\]
If we denote the detected communities as $\wh{\cal N}_1, \ldots, \wh{\cal N}_k$, then 
\beq \label{eq:blindcd} 
F( \wh{\cal N}_1,\ldots, \wh{\cal N}_k ; {\bm U}_k ) - F^\star = {\cal O} ( \eta_k + \sigma + {m}^{-1/2} ).
\eeq
In other words, the BlindCD approaches the performance of SC if the graph filter is $k$-low-pass with $\eta_k \ll 1$, the observation noise $\sigma$ is small, and the number of samples $m$ is large. 
Notice that \eqref{eq:blindcd} is a general result which holds even if $\EE[{\bm x}_\ell {\bm x}_\ell^\top]$ is non-diagonal or low-rank. Moreover, BlindCD is shown to outperform a two-step approach that learns the graph first and then apply SC; see \cite{wai2019blind}.

In Fig. \ref{fig:senate_stock}, we illustrate results of community detection for opinion dynamics and financial data by using data from US Senate from the 115th US Congress (2017-2019) and daily return data of stocks in the S\&P100 index from Feb.~2013 to Dec.~2016 [source: \url{https: //www.kaggle.com/camnugent/sandp500}] respectively. The observed steady-state graph signals ${\bm y}_\ell$ for the opinion dynamics case are the aggregated vote records of each state, and we observe $m=502$ voting rounds. In Fig.~\ref{fig:senate_stock} {\sf(a)}, we apply BlindCD to partition the states into $K=2$ groups, where a close alignment between our results with the actual party memberships of this US Congress is observed.
The financial dataset used contains $m=975$ days of data for $n=92$ stocks. In Fig.~\ref{fig:senate_stock} {\sf (b)} we apply BlindCD to partition the stocks into $K=10$ groups. Each of the  community detected includes companies of the same business type (for instance,  `BAC' (Bank of America) is with `JPM' (JP Morgan)) showing the effectiveness of the method.

\subsection{Smooth Graph Signals}
In Section \ref{sec:basic} we introduced the graph quadratic form to quantify the \emph{smoothness} of a graph signal.
Particularly, if ${\rm S}_{2}( {\bm y} ) = {\bm y}^\top {\bm L} {\bm y} \ll \| {\bm y} \|_2$, the graph signal ${\bm y}$ is said to be smooth.
For $k$-low-pass graph signals, we observe that 
\beq \label{eq:smooth_gs} 
\EE\big[ {\rm S}_{2}( {\bm y}_\ell ) \big] \approx \sum_{i=1}^k \lambda_i \, | h(\lambda_i) |^2 + \sigma^2 {\rm Tr}( {\bm L} ),
\eeq
where we have used that ${\cal H}({\bm L})$ is $k$-low-pass with $\eta_k \ll 1$ to derive the approximations.
In the cases when $\lambda_i \approx 0$, $i=1,\ldots,k$ such as large SBM-PPM graphs with parameters $(a,b)$ satisfying $b \ll 1$, $a \approx 1$, we expect the $k$-low-pass graph signal to be smooth, i.e., $\EE [ {\rm S}_{2}( {\bm y}_\ell ) ] \approx 0$.

\paragraph*{b-i). Graph Topology Learning} The smoothness property can be used to learn the graph topology by fitting a Laplacian matrix which best \emph{smoothens} the graph signals. This is exemplified by the estimator:  
\beq \label{eq:dong_smooth}
\begin{array}{rl}
\ds \min_{ {\bm z}_\ell, \ell=1,\ldots,m, \wh{\bm L} } & \ds \frac{1}{m} \sum_{\ell=1}^m \Big\{ \frac{1}{\sigma^2} \| {\bm z}_\ell - {\bm y}_\ell \|_2^2 + {\bm z}_\ell^\top \wh{\bm L} {\bm z}_\ell \Big\}   \\
\text{s.t.} & {\rm Tr}( \wh{\bm L} ) = n,~\wh{L}_{ji} = \wh{L}_{ij} \leq 0,~i \neq j,~\wh{\bm L}{\bf 1} = {\bm 0},
\end{array}
\eeq
where we have used the graph quadratic form, ${\bm z}_\ell^\top \wh{\bm L} {\bm z}_\ell$, to regulate the smoothness of ${\bm z}_\ell \approx {\bm y}_\ell$ with respect to the fitted $\wh{\bm L}$.  Dong et al.~\cite{dong2016learning} motivated \eqref{eq:dong_smooth}  as a maximum-a-posterior (MAP) estimator for the Laplacian matrix, where ${\bm y}_\ell \sim {\cal N}( {\bm 0} , {\bm L}^\dagger + \sigma^2 {\bm I} )$ and ${\bm L}^\dagger$ is the pseudo-inverse of the Laplacian matrix. This amounts to interpreting the data as outcomes of a Gaussian Markov Random Field (GMRF) with precision matrix chosen as the Laplacian, effectively connecting statistical graphical models to GSP models. 
Note that methods following similar insights as \eqref{eq:dong_smooth} can be found in \cite{kalofolias2016learn,pasdeloup2017characterization}.  

For graph signals that are output from low-pass graph temporal filters, a similar smoothness property to \eqref{eq:smooth_gs} can also be exploited to interpolate missing data. Let $\bm{Y} \in \RR^{n \times m}$ be  
a matrix  whose columns are $\bm{y}_t, t = 1,2,\dots m$, where ${\bm y}_t$ is the graph signal at time $t$; and ${\bm Y}^{\sf samp}$ be the sampled version of ${\bm Y}$ where some values are missing at different node/time indices. 
The key for interpolating the data is to regularize via graph quadratic form and $\ell_2$ norm of the time derivative in addition to minimizing the $\ell_2$ misfit between available samples $\bm{Y}^{\sf samp}$ and reconstructed samples at known locations, $ \mathcal{M}(\bm{Y})$, i.e.,
\beq \notag
\begin{split}
\min_{\bm{Y} \in \RR^{n \times m}} & \norm{\mathcal{M}(\bm{Y})- \bm{Y}^{\sf samp}}_F^2  \\
& + \gamma \, \Big\{ \sum_{\ell=1}^{m} {\bm y}_\ell^\top {\bm L} {\bm y}_\ell + \sum_{\ell=2}^{m} \| {\bm y}_\ell - {\bm y}_{\ell-1} \|_2^2 \Big\}. 
\end{split}
\eeq
See \cite{grassi2017time} and the references therein for a detailed discussion.  
\subsection{Anomalies Detection with Low-pass GSP}
Consider a model consistent with \eqref{eq:lpf_graphsignal}. The fact that the low-pass graph process is dominated by low graph frequency components can be considered the null-hypothesis characterized by the low-pass properties, such as low-rank covariance matrix and smoothness. On the other hand, many anomalies can be modeled as an additive  \emph{sparse noise} signal ${\bm w}_i$, or a \emph{high frequency} graph signal. 
Such noise signals arise in several scenarios such as  a change in the graph connectivity or parameters,  a contingency in infrastructures, the result of malicious activities in social networks, or the sudden fall in the market value of a financial entity. High frequency noise signals are  also produced by a perturbation that is inconsistent with the generative model. For instance, in infrastructure networks this could be symptomatic of malfunctioning sensors or even a false data injection attack (FDIA)\cite{ramakrishna2019detection}. 

Such anomalies cause a surge in the high frequency spectral components of a low-pass graph signal, a fact that can be leveraged in a  manner similar to the classical array processing problem of detecting a source embedded in noise. 
\begin{figure*}
    \centering \begin{tabular}{r c r c}
    \addvbuffer[0cm 4.75cm]{\sf (a)}\hspace{-.45cm}
    & \includegraphics[width=0.3\textwidth]{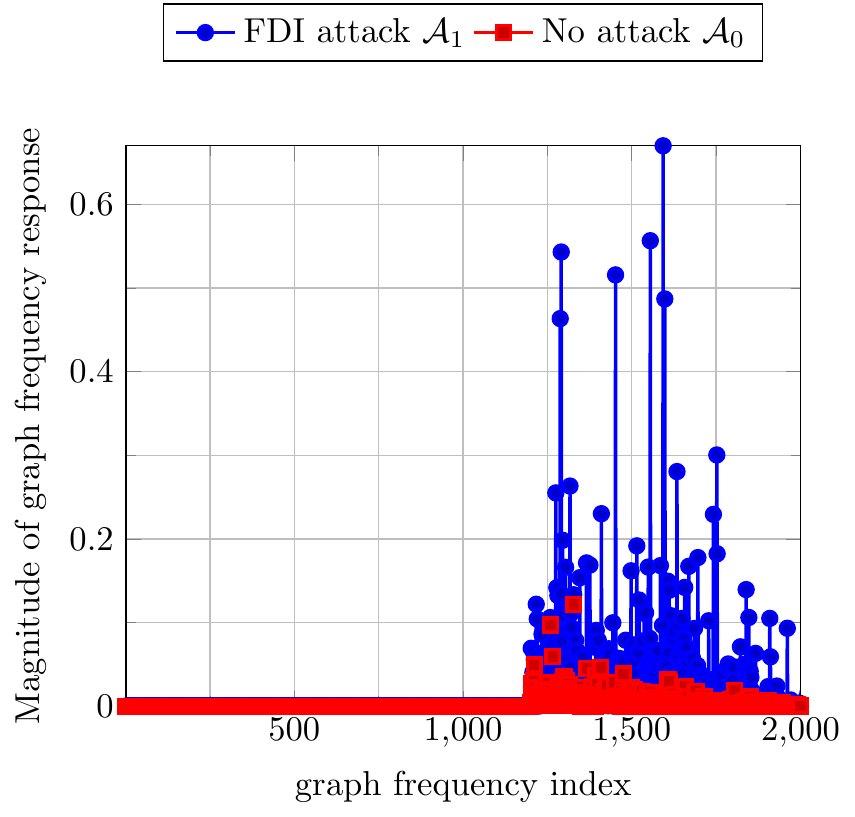}
    & \addvbuffer[0cm 4.75cm]{\sf (b)}\hspace{-.45cm}
    & \includegraphics[width=0.325\textwidth]{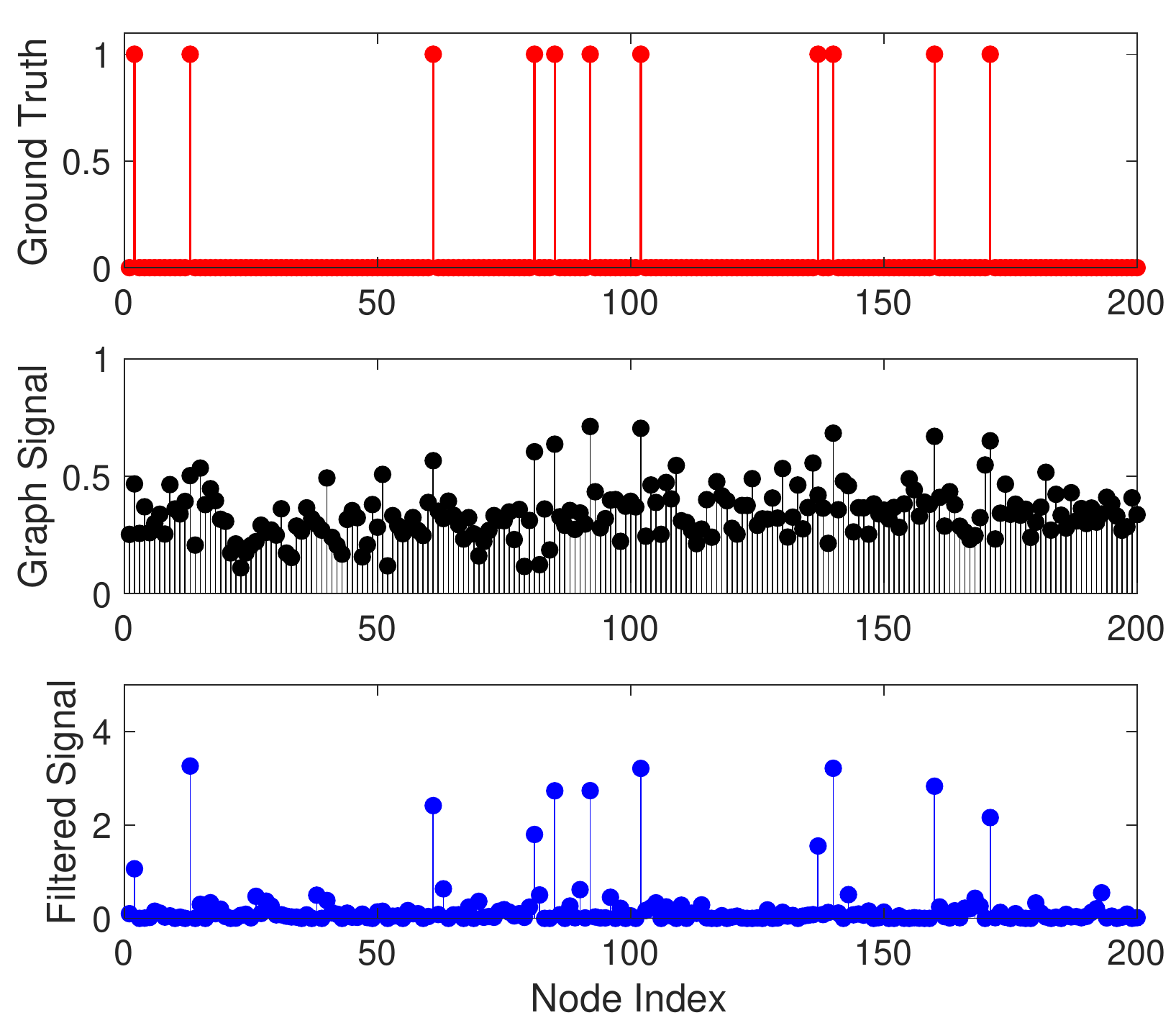}
    \end{tabular}\vspace{-.2cm}
    \caption{({\sf a}) Magnitude of graph Fourier transform of the output after ideal high-pass filtering,  $\abs{\bm{U}^{T}{\cal H}_{\sf HPF} ({\bm L})\bm{y}_{\ell}}$, $k=1200$, under the hypotheses of anomaly $\mathcal{A}_{1}$ and no anomaly $\mathcal{A}_{0}$. 
    Particular example of FDIA on voltage graph signal $ \bm{y}_{\ell}$ from ACTIVSg2000 case is shown. 
    We see a surge in high frequency components when there is an attack. ({\sf b}) Spatial difference filtering of the graph signal under a diffusion model with abnormal activities on 11 nodes; the top figure shows the ground truth locations of anomalies, middle and bottom figures show the graph signal $\bm{y}_{\ell} $ and filtered graph signal,  ${\cal H}_{\sf SD}( {\bm L})\bm{y}_{\ell} $, respectively.\vspace{-.2cm}}
    \label{fig:anomaly_detection_FDI}
\end{figure*}
Formally, the observed signal under null and alternative hypothesis is described as 
\beq \notag
{\bm y}_\ell = \begin{cases}
{\cal H}({\bm L}) {\bm x}_\ell, & ~~\text{under}~{\cal A}_0, \\
{\cal H}({\bm L}) {\bm x}_\ell + {\bm w}_\ell &~~ \text{under}~{\cal A}_1,
\end{cases}
\eeq
where ${\bm w}_\ell$ is a high frequency graph signal. 
Our task amounts to testing the hypothesis ${\cal A}_0$, ${\cal A}_1$, and/or to estimate the locations of non-zeros in ${\bm w}_i$ when the latter is a sparse signal and under ${\cal A}_1$. 

Intuitively, we can apply a high-pass graph filter to distinguish between ${\cal A}_0$ and ${\cal A}_1$. 
Let ${\cal H}_{\sf HPF} ({\bm L})$ be an ideal high-pass graph filter with the frequency response $h_{\sf HPF}({\lambda}) = 1$, $\lambda \geq \lambda_{k+1}$ and $0$ otherwise. Consider the test statistics as $\Gamma_\ell = \| {\cal H}_{\sf HPF}( {\bm L} ) {\bm y}_\ell \|$.
Under ${\cal A}_0$ and the $k$-low-pass assumption, we have $\Gamma_\ell \leq \| {\cal H}_{\sf HPF}( {\bm L} ) {\cal H}({\bm L}) \|_2 \| {\bm x}_\ell \| = \| h_{\sf HPF}(\bm{\lambda}) \odot h(\bm{\lambda}) \|_\infty \| {\bm x}_\ell \| = {\cal O}(\eta_k)$, and thus the test statistics $\Gamma_\ell$ will be small. On the other hand, under ${\cal A}_1$, we obtain ${\cal H}_{\sf HPF}( {\bm L} ) {\bm y}_\ell \approx {\bm w}_\ell$ since the anomalies consist of high graph frequency components. Thus the test statistics $\Gamma_\ell$ will be large. 
Imposing a threshold of $\delta = \Theta( \eta_k )$, we can consider the detector
\[
\Gamma_\ell = \| {\cal H}_{\sf HPF}( {\bm L} ) {\bm y}_\ell \| \underset{{\cal A}_1}{\overset{{\cal A}_0}{\lessgtr}} \delta.
\]
Furthermore, if ${\cal A}_1$ holds, these anomaly events can be {located} from the support of ${\cal H}_{\sf HPF}( {\bm L} ) {\bm y}_\ell$. 

As a demonstration, Fig.~\ref{fig:anomaly_detection_FDI} (a) shows the magnitude of GFT of the voltage graph signal after filtering using an ideal high-pass graph filter, $\bm{U}^\top {\cal H}_{\sf HPF}( {\bm L} ) {\bm y}_\ell $. The voltage graph signal under the hypothesis of no anomaly is the output of a low-pass graph filter. When there is a FDIA, we observe an increase in energy of the high frequency components. 

To obtain a simple implementation of high-pass graph filters, we may consider ${\cal H}_{\sf SD}( {\bm L} ) = {\bm L} = {\bm D} - {\bm A}$ whose frequency response is given by $h({\lambda}) = {\lambda}$. When applied on a graph signal ${\bm y}_\ell$, we will observe the difference between ${\bm D} {\bm y}_\ell$ and ${\bm A} {\bm y}_\ell$, where the latter is a one-hop averaged version of ${\bm y}_\ell$. We call this operation the spatial difference which is similar to the method proposed in \cite{wai2018identifying} for anomaly detection on social networks. See the illustration in Fig.~\ref{fig:anomaly_detection_FDI} (b).  

\section{Concluding Remarks}\label{sec:conc}
In this user guide, we highlighted the key elements of low-pass GSP in several applications like graph parameter inference and graph signal sampling while emphasizing the intuition from time series analysis. We also discussed several physical models where low-pass GSP can be effectively used. 
However, the tools available for low-pass  GSP are ever-expanding, and aid the discovery of new physical models where low-pass GSP can be applied. Additionally, there are several open research directions as discussed below.

\paragraph{Directed Graphs}
Throughout this article, we have assumed that the observed data is supported on a graph topology which is undirected, and the shift operator (Laplacian matrix) is symmetric. This is clearly not a truthful model for a lot of real systems such as social and economics networks. The challenge of extending the existing GSP tools to directed graphs lies in defining the appropriate GFT basis. 
For instance, the properties of a circular shift matrix is what a directed shift operator should emulate. 

Much of the prior research has focused on finding the appropriate GFT basis on directed graphs.
The definition of frequency is again variational, but based on the norm of the difference between and vector and the shift operator $\bm S$ of the corresponding graph  does not have to be symmetric. More formally,  the idea of smoothness is defined as $\|\bm x -\lambda_n^{-1}\bm S \bm x\|_2^2$
where $\lambda_n$ is the maximum eigenvalue of $\bm S$. This is the definition used in  \cite{sandryhaila2013discrete} for GFT on directed graphs, where the GSO is set as the adjacency matrix $\bm A$ and the GFT is defined as $\wt{\bm y} = {\bf U}^{-1} {\bm y}$ such that ${\bf U}$ is obtained from the Jordan decomposition of the adjacency matrix ${\bm A} = {\bf U} \bm{\Lambda} {\bf U}^{-1}$. 
Although ${\bf U}$ is a basis, it is not orthogonal, so the Parseval's identity does not hold since $\| \wt{\bm y} \|_2 \neq \| {\bm y} \|_2$. That is not surprising since the norm of  $\wt{\bm y}$ does not have the same physical interpretation of power spectral density that applies to signals whose support is time. A potential fix is studied in \cite{sardellitti2017graph} which searches for the GFT basis that minimizes the directed total variation, also see \cite{shafipour2018directed}. Unfortunately the GFT basis does not admit a closed form solution. 

The tools discussed in this article, such as sampling theory \cite{chen2015discrete} and anomaly detection may still work for low-pass graph signals on directed graphs with minor adjustments. 
The challenge lies in the graph inference/learning methods since second-order statistics such as correlations are difficult to justify in directed graphs, where also the notion of \emph{community} is ambiguous. A useful definition of community must first be studied before tools of GSP can be applied for community inference in directed graphs.


\paragraph{Low-pass Graph Signals in the Edge Space}
An alternative form of graph signals are those that are defined on the edges. They can be defined as the function $f: {\cal E} \rightarrow \RR$ and the equivalent vector ${\bm f} \in \RR^{|{\cal E}|}$, which are useful for describing \emph{flows} on graphs such as traffic in transportation network. As suggested in \cite{schaub2018flow}, the shift operator can be taken as the edge Laplacian ${\bf L}_e = {\bf B}^\top {\bf B}$, where ${\bf B} \in \RR^{n \times |{\cal E}|}$ is the node-to-edge incidence matrix. The null space of the edge Laplacian ${\bf L}_e$ corresponds to the cyclic flow vector, which is also the eigenvector for the lowest graph frequency $\lambda_1 = 0$. It is anticipated that a low-pass edge-graph signal, whose energy is focused in the low graph frequencies, will consist of mostly cyclic flows within communities.
An interesting direction is to develop a sampling theory for low-pass edge-graph signals, as well as the inference of edge Laplacian matrix. 

\paragraph{Identifying Low-pass Graph Signals}
So far we have relied on domain knowledge about the data models, e.g., the examples in Section~\ref{sec:model}, to help justify various graph data as low-pass graph signals. 

For graph signals taken from an unknown system, one has to be cautious before applying this low-pass GSP user guide. Even though \emph{non-low-pass} graph processes are rarely found in a natural setting, there is a crucial need to design tools for identifying low-pass graph signals. With known GSO, it can be done by inspecting the GFT spectrum; with unknown GSO, the problem is related to the joint estimation of graph process and topology; readers are referred to \cite{ioannidis2019semi} for recent works in this direction.

\paragraph*{Acknowledgement} The authors thank the anonymous reviewers and the guest editors for their useful feedback.
This material is based upon work supported in part by, the U. S. Army Research Laboratory and the U. S. Army Research Office under contract/grant number W911NF2010153 and the NSF CCF-BSF: CIF: 1714672 grant. Hoi-To Wai's work is supported by the CUHK Direct Grant \#4055135.

\linespread{1.1}
\normalsize

\bibliographystyle{IEEEtran}
\bibliography{ref_list_graph}

\begin{thebibliography}{10}
\providecommand{\url}[1]{#1}
\csname url@samestyle\endcsname
\providecommand{\newblock}{\relax}
\providecommand{\bibinfo}[2]{#2}
\providecommand{\BIBentrySTDinterwordspacing}{\spaceskip=0pt\relax}
\providecommand{\BIBentryALTinterwordstretchfactor}{4}
\providecommand{\BIBentryALTinterwordspacing}{\spaceskip=\fontdimen2\font plus
\BIBentryALTinterwordstretchfactor\fontdimen3\font minus
  \fontdimen4\font\relax}
\providecommand{\BIBforeignlanguage}[2]{{%
\expandafter\ifx\csname l@#1\endcsname\relax
\typeout{** WARNING: IEEEtran.bst: No hyphenation pattern has been}%
\typeout{** loaded for the language `#1'. Using the pattern for}%
\typeout{** the default language instead.}%
\else
\language=\csname l@#1\endcsname
\fi
#2}}
\providecommand{\BIBdecl}{\relax}
\BIBdecl

\bibitem{kolaczyk2014statistical}
E.~D. Kolaczyk and G.~Cs{\'a}rdi, \emph{{Statistical analysis of network data
  with R}}.\hskip 1em plus 0.5em minus 0.4em\relax Springer, 2014, vol.~65.

\bibitem{shuman2013emerging}
D.~I. Shuman, S.~K. Narang, P.~Frossard, A.~Ortega, and P.~Vandergheynst,
  ``{The emerging field of signal processing on graphs: Extending
  high-dimensional data analysis to networks and other irregular domains},''
  \emph{IEEE Signal Processing Magazine}, vol.~30, no.~3, pp. 83--98, 2013.

\bibitem{ortega2018graph}
A.~Ortega, P.~Frossard, J.~Kova{\v{c}}evi{\'c}, J.~M. Moura, and
  P.~Vandergheynst, ``{Graph signal processing: Overview, Challenges, and
  Applications},'' \emph{Proceedings of the IEEE}, vol. 106, no.~5, pp.
  808--828, 2018.

\bibitem{sandryhaila2013discrete}
A.~Sandryhaila and J.~M. Moura, ``{Discrete Signal Processing on Graphs},''
  \emph{IEEE Transactions on Signal Processing}, vol.~61, no.~7, pp.
  1644--1656, 2013.

\bibitem{tremblay2018design}
N.~Tremblay, P.~Gon{\c{c}}alves, and P.~Borgnat, ``{Design of Graph Filters and
  Filterbanks},'' in \emph{Cooperative and Graph Signal Processing}.\hskip 1em
  plus 0.5em minus 0.4em\relax Elsevier, 2018, pp. 299--324.

\bibitem{wai2019blind}
H.-T. Wai, S.~Segarra, A.~E. Ozdaglar, A.~Scaglione, and A.~Jadbabaie, ``{Blind
  Community Detection From Low-Rank Excitations of a Graph Filter},''
  \emph{IEEE Transactions on Signal Processing}, vol.~68, pp. {436--451}, 2020.

\bibitem{rohe2011spectral}
K.~Rohe, S.~Chatterjee, B.~Yu \emph{et~al.}, ``Spectral clustering and the
  high-dimensional stochastic blockmodel,'' \emph{The Annals of Statistics},
  vol.~39, no.~4, pp. 1878--1915, 2011.

\bibitem{ding2010spectral}
X.~Ding, T.~Jiang \emph{et~al.}, ``{Spectral distributions of adjacency and
  Laplacian matrices of random graphs},'' \emph{The Annals of Applied
  Probability}, vol.~20, no.~6, pp. 2086--2117, 2010.

\bibitem{isufi2016separable}
E.~Isufi, A.~Loukas, A.~Simonetto, and G.~Leus, ``Separable autoregressive
  moving average graph-temporal filters,'' in \emph{2016 24th European Signal
  Processing Conference (EUSIPCO)}.\hskip 1em plus 0.5em minus 0.4em\relax
  IEEE, 2016, pp. 200--204.

\bibitem{thanou2017learning}
D.~Thanou, X.~Dong, D.~Kressner, and P.~Frossard, ``{Learning Heat Diffusion
  Graphs},'' \emph{IEEE Transactions on Signal and Information Processing over
  Networks}, vol.~3, no.~3, pp. 484--499, 2017.

\bibitem{ravazzi2017learning}
C.~Ravazzi, R.~Tempo, and F.~Dabbene, ``{Learning Influence Structure in Sparse
  Social Networks},'' \emph{IEEE Transactions on Control of Network Systems},
  vol.~5, no.~4, pp. 1976--1986, 2017.

\bibitem{friedkin2011formal}
N.~E. Friedkin, ``{A Formal Theory of Reflected Appraisals in the Evolution of
  Power},'' \emph{Administrative Science Quarterly}, vol.~56, no.~4, pp.
  501--529, 2011.

\bibitem{candogan2012optimal}
O.~Candogan, K.~Bimpikis, and A.~Ozdaglar, ``{Optimal Pricing in Networks with
  Externalities},'' \emph{Operations Research}, vol.~60, no.~4, pp. 883--905,
  2012.

\bibitem{billio2012econometric}
M.~Billio, M.~Getmansky, A.~W. Lo, and L.~Pelizzon, ``Econometric measures of
  connectedness and systemic risk in the finance and insurance sectors,''
  \emph{Journal of financial economics}, vol. 104, no.~3, pp. 535--559, 2012.

\bibitem{mantegna1999introduction}
R.~N. Mantegna and H.~E. Stanley, \emph{Introduction to econophysics:
  correlations and complexity in finance}.\hskip 1em plus 0.5em minus
  0.4em\relax Cambridge university press, 1999.

\bibitem{glover2008power}
J.~D. Glover, M.~Sarma, and T.~Overbye, ``{Power System Analysis and Design},''
  \emph{Cengage Learning}, vol.~4, 2008.

\bibitem{ramakrishna2019modeling}
R.~Ramakrishna and A.~Scaglione, ``{On Modeling Voltage Phasor Measurements as
  Graph Signals},'' in \emph{2019 IEEE Data Science Workshop, DSW 2019}.\hskip
  1em plus 0.5em minus 0.4em\relax Institute of Electrical and Electronics
  Engineers Inc., 2019, pp. 275--279.

\bibitem{chen2015discrete}
S.~Chen, R.~Varma, A.~Sandryhaila, and J.~Kova{\v{c}}evi{\'c}, ``{Discrete
  Signal Processing on Graphs: Sampling Theory},'' \emph{IEEE Transactions on
  Signal Processing}, vol.~63, no.~24, pp. 6510--6523, 2015.

\bibitem{anis2016efficient}
A.~Anis, A.~Gadde, and A.~Ortega, ``{Efficient Sampling Set Selection for
  Bandlimited Graph Signals Using Graph Spectral Proxies},'' \emph{IEEE
  Transactions on Signal Processing}, vol.~64, no.~14, pp. 3775--3789, 2016.

\bibitem{tsitsvero2016signals}
M.~Tsitsvero, S.~Barbarossa, and P.~Di~Lorenzo, ``{Signals on graphs:
  Uncertainty Principle and Sampling},'' \emph{IEEE Transactions on Signal
  Processing}, vol.~64, no.~18, pp. 4845--4860, 2016.

\bibitem{dong2016learning}
X.~Dong, D.~Thanou, P.~Frossard, and P.~Vandergheynst, ``{Learning Laplacian
  Matrix in Smooth Graph Signal Representations},'' \emph{IEEE Transactions on
  Signal Processing}, vol.~64, no.~23, pp. 6160--6173, 2016.

\bibitem{kalofolias2016learn}
V.~Kalofolias, ``{How to Learn a Graph from Smooth Signals},'' in
  \emph{Artificial Intelligence and Statistics}, 2016, pp. 920--929.

\bibitem{pasdeloup2017characterization}
B.~Pasdeloup, V.~Gripon, G.~Mercier, D.~Pastor, and M.~G. Rabbat,
  ``{Characterization and Inference of Graph Diffusion Processes From
  Observations of Stationary Signals},'' \emph{IEEE Transactions on Signal and
  Information Processing over Networks}, vol.~4, no.~3, pp. 481--496, 2017.

\bibitem{grassi2017time}
F.~Grassi, A.~Loukas, N.~Perraudin, and B.~Ricaud, ``{A Time-Vertex Signal
  Processing Framework: Scalable Processing and Meaningful Representations for
  Time-Series on Graphs},'' \emph{IEEE Transactions on Signal Processing},
  vol.~66, no.~3, pp. 817--829, 2017.

\bibitem{ramakrishna2019detection}
R.~Ramakrishna and A.~Scaglione, ``{Detection of False Data Injection Attack
  using Graph Signal Processing for the Power Grid},'' in \emph{{2019 IEEE
  Global Conference on Signal and Information Processing (GlobalSIP)}}.\hskip
  1em plus 0.5em minus 0.4em\relax Institute of Electrical and Electronics
  Engineers Inc., 2019, pp. {1--5}.

\bibitem{wai2018identifying}
H.-T. Wai, A.~E. Ozdaglar, and A.~Scaglione, ``{Identifying Susceptible Agents
  in Time Varying Opinion Dynamics Through Compressive Measurements},'' in
  \emph{{2018 IEEE International Conference on Acoustics, Speech and Signal
  Processing (ICASSP)}}.\hskip 1em plus 0.5em minus 0.4em\relax IEEE, 2018, pp.
  4114--4118.

\bibitem{sardellitti2017graph}
S.~Sardellitti, S.~Barbarossa, and P.~Di~Lorenzo, ``{On the Graph Fourier
  Transform for Directed Graphs},'' \emph{IEEE Journal of Selected Topics in
  Signal Processing}, vol.~11, no.~6, pp. 796--811, 2017.

\bibitem{shafipour2018directed}
R.~Shafipour, A.~Khodabakhsh, G.~Mateos, and E.~Nikolova, ``{A Directed Graph
  Fourier Transform With Spread Frequency Components},'' \emph{IEEE
  Transactions on Signal Processing}, vol.~67, no.~4, pp. 946--960, 2018.

\bibitem{schaub2018flow}
M.~T. Schaub and S.~Segarra, ``Flow smoothing and denoising: graph signal
  processing in the edge-space,'' in \emph{2018 IEEE Global Conference on
  Signal and Information Processing (GlobalSIP)}.\hskip 1em plus 0.5em minus
  0.4em\relax IEEE, 2018, pp. 735--739.

\bibitem{ioannidis2019semi}
V.~N. Ioannidis, Y.~Shen, and G.~B. Giannakis, ``{Semi-Blind Inference of
  Topologies and Dynamical Processes Over Dynamic Graphs},'' \emph{IEEE
  Transactions on Signal Processing}, vol.~67, no.~9, pp. 2263--2274, 2019.

\end{thebibliography}
 
\end{document}